\documentclass[pdflatex, sn-aps]{sn-jnl}

\usepackage{amsmath,amssymb,amsfonts,kpfonts, amsthm}
\usepackage[title]{appendix}
\usepackage{bookmark}
\usepackage{booktabs}
\usepackage[T1]{fontenc}
\usepackage{graphicx}
\usepackage{manyfoot}
\usepackage{textcomp}
\usepackage[table,svgnames]{xcolor}

\usepackage{algorithm}
\usepackage{algorithmicx}
\usepackage{algpseudocode}
\usepackage{listings}

\usepackage{multirow}
\usepackage{multicol}
\usepackage{transparent}
\usepackage{tablefootnote}

\usepackage{hyperref}

\usepackage{silence}
\usepackage{caption}
\usepackage{subcaption}

\newtheoremstyle{spacedef}
  {1em}   
  {1em}   
  {\itshape} 
  {}      
  {\bfseries} 
  {.}     
  { }     
  {}      
\theoremstyle{spacedef}
\newtheorem{theorem}{Definition}[section]

\newcommand{\first}{\textcolor{Green}}{}
\newcommand{\second}{\textcolor{YellowGreen}}{}
\newcommand{\third}{\textcolor{Olive}}{}

\definecolor{lastcolor}{RGB}{204,0,0}
\definecolor{penultimatecolor}{RGB}{255,102,0}
\definecolor{thirdtolastcolor}{RGB}{255,153,0}
\newcommand{\last}{\textcolor{lastcolor}}{} 
\newcommand{\penultimate}{\textcolor{penultimatecolor}}{} 
\newcommand{\thirdtolast}{\textcolor{thirdtolastcolor}}{} 

\begin{document}

\title{Rank-Refining Seed Selection Methods for Budget Constrained Influence Maximisation in Multilayer Networks under Linear Threshold Model}

\author*[1]{\fnm{Micha{\l}} \sur{Czuba}}\email{michal.czuba@pwr.edu.pl}

\author[1]{\fnm{Piotr} \sur{Br\'{o}dka}}\email{piotr.brodka@pwr.edu.pl}

\affil[1]{\orgdiv{Department of Artificial Intelligence}, \orgname{Wrocław University of Science and Technology}, \orgaddress{\street{27 Wybrze\.{z}e Wyspia\'{n}skiego}, \city{Wrocław}, \postcode{50-370}, \country{Poland}}}

\abstract{The problem of selecting an optimal seed set to maximise influence in networks has been a subject of intense research in recent years. However, despite numerous works addressing this area, it remains a topic that requires further elaboration. Most often, it is considered within the scope of classically defined graphs with a spreading model in the form of Independent Cascades. In this work, we focus on the problem of budget-constrained influence maximisation in multilayer networks using a Linear Threshold Model. Both the graph model and the spreading process we employ are less prevalent in the literature, even though their application allows for a more precise representation of the opinion dynamics in social networks. This paper aims to answer which of the sixteen evaluated seed selection methods is the most effective and how similar they are. Additionally, we focus our analysis on the impact of spreading model parameters, network characteristics, a budget, and the seed selection methods on the diffusion effectiveness in multilayer networks. Our contribution also includes extending several centrality measures and heuristics to the case of such graphs. The results indicate that all the factors mentioned above collectively contribute to the effectiveness of influence maximisation. Moreover, there is no seed selection method which always provides the best results. However, the seeds chosen with VoteRank-based methods (especially with the $v-rnk-m$ variant we propose) usually provide the most extensive diffusion.}


\maketitle

\section{Introduction}\label{sec:intro}

Influence maximisation in social networks aims to identify a set of initial nodes (seeds) that will maximally spread influence throughout the network. Since the initial work of Kempe~\cite{kempe2003maximizing} on maximising the spread of influence in social networks, researchers all over the world have been working on inventing and perfecting methods that will produce the best seed set in terms of overall network influence.

\begin{figure}[h]
    \centering
    \includegraphics[width=.7\linewidth]{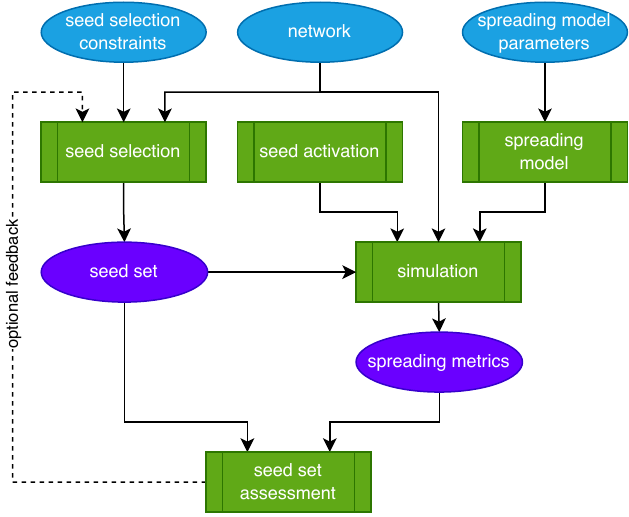}
    \caption{Schematic depiction of the influence maximisation's components. Light blue ellipses --- input data, dark blue ellipses --- output data, green rectangles --- functions, arrows --- order of execution.}
    \label{fig:im_framework}
\end{figure}

The influence maximisation problem depends on several factors (see Fig.~\ref{fig:im_framework}), which ultimately translate into the dynamics of influence spread in the network and, consequently, the final number of activated agents. We can distinguish here the network itself (e.g. directed, static, temporal, multilayer, hypergraph), the spreading model (e.g. Independent Cascade, Q-Voter, Suspected-Infected-Recovered), its parameters (e.g. $\beta$ in Suspected-Infected-Recovered), the method of selecting seeds (e.g. random, greedy, degree-based ranking), the way seeds are used (all at once, sequentially during the propagation, etc.), constraints on the seed selection (e.g. budget, cost, time), and property of the influence to be optimised (for instance the final number of activated agents, the number of simulation steps needed to activate $50\%$ of agents in the graph). Moreover, these factors are usually interdependent (as denoted with the feedback loop in Fig.~\ref{fig:im_framework}). For instance, seed set selection can be closely related to the spreading model, which is very common in methods utilising the optimisation mechanism. As one can note, the number of degrees of freedom of the problem is considerable --- the problem of influence maximisation belongs to the NP-hard group~\cite{kempe2003maximizing}. Furthermore, many of them are interdependent (for instance, the network type and method of selecting seeds). That is why there have been so many studies in the field of influence maximisation, and researchers still address this problem. Specifying all these factors allows for a precise definition of the problem. Below, we refer to them according to the title of our work to show how the problem we tackle is positioned in the literature.

\subsection{Seeding process}

Research in previous years led to the development of multiple seed selection strategies (methods), which can be divided into three categories: Simulation-based, Heuristic-based, and Mixed Approaches, each differing in their problem-solving perspective.

\subsubsection{Seed Selection Strategy}

Simulation-based approaches utilise Monte Carlo (MC) simulations to compute the spread of influence in a network, which can be computationally intensive. These methods perform numerous explicit MC simulations to identify nodes with the highest marginal gain. However, methods belonging to this group may not be computationally efficient for large-scale networks due to the high number of MC simulations required. Researchers have therefore focused on optimising them by either reducing the number of MC simulations needed to find a seed set or the complexity of each MC simulation~\cite{singh2022influence}. The classic greedy algorithm proposed in~\cite{kempe2003maximizing} and its extensions such as CELF~\cite{leskovec2007cost} or CGA~\cite{wang2010communitybasedgreedy} are examples of methods belonging to this category.

Heuristic-based approaches utilise approximate scoring methods to evaluate the potential influence of each node without relying on time-consuming MC simulations. These methods are often more scalable and efficient compared to simulation-based approaches. Heuristic approaches typically leverage features like network topology or corresponding diffusion model to determine the seed set. They can be categorised into two main groups: rank refinement and model reduction~\cite{singh2022influence}. Rank refinement methods assign a rank to each agent based on computed metrics for estimating the influence spread. The seed set is then determined based on these scores. Examples of such methods include Degree Centrality, Betweenness Centrality, Closeness Centrality, and more sophisticated methods like Single Discount method~\cite{sheikhahmadi2015improving}, Degree Punishment method~\cite{wang2016effective}, VoteRank~\cite{zhang2016identifying}, community-based methods~\cite{he2015novel}, clustering-based methods~\cite{bao2017identifying}, network control-based methods~\cite{sadaf2022maximising} and many more. Model reduction methods simplify the computation of expected influence spread by reducing the complexity of the diffusion model. This can be achieved either by reducing stochastic models to deterministic models or by restricting the influence on the local region of the network. Approaches such as Shortest Path Model~\cite{kimura2006tracable}, Local Directed Acyclic Graph heuristic introduced in~\cite{cheng2010ScalableIM} or the Independent Path algorithm~\cite{kim2013scalable} are representatives of this category.

According to the taxonomy introduced by~\cite{singh2022influence}, mixed approaches aim to enhance the theoretical efficiency of simulation-based methods while ensuring a guarantee of approximation. These approaches precompute a set of sketches based on a specific diffusion model to avoid the need for rerunning MC simulations. Subsequently, they leverage these sketches to compute the spread of influence. Mixed approaches were categorised into two classes based on the way the sketches are generated: snapshot-based (e.g. NewGreedy~\cite{cheng2013staticgreedy} or StaticGreedy~\cite{zhou2015upper}) and Reverse Reachable Sets~\cite{borgs2014reversereachablesets}, e.g.~\cite{Sun2021InfMaxRR}. Regarding this kind of division, one can append to this group methods based on artificial intelligence (classic machine learning, deep learning, etc), for instance~\cite{tiukhova2022influencer, kou2023identify, hajarathaiah2022generalization, rezaei2023machine, bucur2020top} or~\cite{chen2023deepim}.

\subsubsection{Seed Activation Strategy}\label{subsubsec:seed_activation_strategy}

Another aspect of the seeding process is the seed activation strategy~\cite{brodka2021sequential}. Most research focuses on the scenario where all seeds are activated at the same time, but in recent years, researchers started to ask the question, what if we activate only a few seeds at the beginning and see what happens with the spreading process? This led to several approaches like Sequential Seeding~\cite{jankowski2017balancing}, Adaptive Seeding~\cite{seeman2013adaptive}, Seeding Scheduling~\cite{goldenberg2018timing} or Active Seeding~\cite{sela2018active}, which have a different name but the same mechanism behind it. In a nutshell, at the beginning, we select a few seeds using some seed selection strategy. Then, instead of activating them all at the beginning, we activate them in batches, e.g., a few of them at the beginning to start spreading and one more in every iteration, or we add the next batch of seeds when the process is slowing down or dying out. This approach allows us to observe the behaviour of the spreading process and avoid seeding nodes that got activated by their neighbours. Saved seeds can then be used to activate new nodes in sections of the network where no nodes were activated. 

\subsubsection{Seed Selection Constraints}

The process of selecting seeds is subject to various constraints that must be carefully considered during the problem definition. The most commonly used limitation is a budget, which restricts the number of seeds that can be chosen. In some works, researchers assume this constraint in an edge form --- they limit it to one agent~\cite{kitsak2010superspreaders}. Yet, considering only a single super-spreader puts the problem in a different frame. First, the choice of the spreading model gets narrowed down. For example, in LTM, selecting a single-element seed set can result in an inability to trigger diffusion for thresholds that do not oscillate around $0\%$ (and therefore, in an overwhelming number of cases). On the other hand --- such constraint in stochastic diffusion models can make it more difficult to assess such a set as the "spreading potential" of the particular agent depends on many stochastic events (e.g. in ICM), and thus, it can attain a big variance. Another border case is a budget size which leads to an instant saturation of the diffusion. Namely, if one set is too big, then even choosing a seed set randomly will result in a decent diffusion. Thus, if one needs to assess various seed selection methods, the budget size needs to reflect a non-trivial spreading regime. In this work, we follow that approach and do not examine the entire range of budgets. Moreover, we apply different spectra to different diffusion conditions (see Sec.~\ref{sec:setup}). Finally, the effect of this constraint on the rank refining methods is also pronounced. For such approaches, it plays a role of the cutoff applied on the ranking. The bigger it is, the more similar the sets obtained with different methods, as shown later in Fig.~\ref{fig:similarities_rankings_toy_net}.

With regard to other forms of constraints, some studies consider the cost associated with each agent~\cite{SVIRIDENKO2004Knapsack}. In addition to the limitation of the seed set cardinality, they restrict the overall price of selecting particular agents, which cost differently, akin to a rucksack problem where the objective is to maximise the total influence under a budgetary limit.

Structural network properties, which arise from the network topology, where certain layers or nodes have specific roles or connectivity patterns that influence their suitability as seeds, can also play a role of the constraints imposed on the seed selection process. For instance, in~\cite{yuan2024gbim}, authors introduce a seed selection method in the environment consisting of a social network and several information items which spread concurrently and influence each other according to the "association graph" (e.g. spread of item $X$ induces spread of item $Y$). By that, they solve Multiple Influence Maximisation problem\footnote{in their work it is called Multiplex Influence Maximisation} (see Sec.~\ref{subsec:infmaxmln}) constrained by selection of both the agents and the information item to start spread with. Other structural constraints can be rooted in the dynamic nature of the network, where nodes or edges might change over time due to events like link failure, node removal, or structural reconfigurations~\cite{michalski2014temporal}. Another important constraint associated with this group is the assumption of network completeness, which predominates in the literature. However, some studies address the challenge of incomplete data~\cite{liu2024rsif_2023_0625}.

In addition, we can also imagine attribute-based constraints, which result from the particular form of the graph, i.e. when considering a problem of sexually transmitted diseases that both sexes can carry but affect mostly one of them, e.g. HPV~\cite{hpv2024usgov}.

\subsection{Spreading Performance Metrics}\label{subsec:spreading_performance}

In the influence maximisation problem, the objective function can also adopt many forms. How one defines it can determine the entire experimental setup and strongly affect the conclusions. According to~\cite{singh2022influence}, metrics to measure the performance of the influence maximisation can be divided into four groups: Quality, Efficiency, Scalability, and Robustness. Each of these groups treats the objective differently.

The most commonly occurring is the first one. The basic form of such metric is the number of active agents counted at the steady state of the simulation (see Def.~\ref{def:seeds}). There are also works which define an objective function from a more particular point of view. For instance, the domain of Profit Maximisation addresses the problem of the distinction between the influence and adoption of the advertised product (or opinion), and it extends classical diffusion models to incorporate these factors~\cite{wei2012profitmaximisationltm}. An exemplary metric, named "profit", that addresses that problem was proposed by~\cite{tang2018profitmaximsation}, and it combines the benefit obtained from the influence spread and the cost of the influence itself. However, despite being dominant in the literature, metrics belonging to this group have some limitations. Since we focus this study on the influence maximisation efficiency, some of them are addressed in the following parts of the article (see Sec.~\ref{subsec:infmax}).

Works belonging to the second group consider the objective in terms of time. The authors of~\cite{singh2022influence} mention its representatives, such as~\cite{tang2015influence}. Nonetheless, they allocate to this group only works that measure the time complexity of the influence maximisation algorithm. However, we believe this group can be extended by metrics based on the diffusion time considered as the simulation steps. Here, the fundamental one can be the number of simulation steps in the experiment. Meanwhile, the more sophisticated approaches take the form of a combination of the number of activated agents. For instance, in works from the domain of epidemiology, the important parameter is the maximal growth of the infected population~\cite{watroba2023influence}. Other works consider metrics like the number of simulation steps needed to cover a particular area of the network~\cite{sadaf2022maximising}.

The last two groups are Scalability and Robustness. The former one evaluates influence maximisation algorithms not only by their running time but also by memory consumption. Next, Robustness in influence maximisation considers the problem of how an alternation of the diffusion function affects the performance of the seed selection methods. Therefore, a robust method is one which yields satisfactory results under various diffusion regimes~\cite{kempe2016robustness}.

\subsection{Influence Maximisation in Multilayer Networks}\label{subsec:infmaxmln}

Despite the wealth of research on finding optimal seed sets or influential nodes (please see the recent review papers~\cite{ou2022identifying, singh2022influence} on that topic), most works are limited to simple one-layer (aka. singe-layer) networks. However, in the real world, we are not influenced just by our family, friends, or office colleagues. We are being affected in all social circles. What is more, if we are convinced by our family (to the brand of the smartphone, the idea, the political party, etc.), we will become convinced (active) in all social spheres in which we function and eventually start affecting our acquittances. In order to capture all those interactions, we need to use the multilayer networks~\cite{salehi2015spreading}.

The research on influence maximisation in multilayer networks is much more limited than on one-layer networks. Firstly, there are more degrees of complexity due to the multilayer network structure. Secondly, methods developed for simple one-layer networks often do not work correctly in the multilayer scenario~\cite{zhao2013identifying, kitsak2010superspreaders, erlandsson2018seed}. Thus, researchers usually need to start by extending existing methods so they work on multilayer networks and then evaluate them to see which yields the best results. Another problem (that is a reason for the unambiguity in the area of influence maximisation in multilayer networks) is how we define the seed --- whether it is a node on a particular layer or an actor represented by various nodes on different layers.

\cite{singh2022influence} distinguishes three groups of frameworks that consider multilayer networks: Influence Maximisation Across Multiple Networks (IM2), Multiple Influence Maximisation (MIM), and Multiple Influence Maximisation Across Multiple Networks (MIM2). In this study, we tackle the first of them. It assumes that there is one process that spreads through a multilayer network. The second group mentioned considers that more than one process affects the graph. Although the multilaterance is not introduced directly there, we can depict that problem as a diffusion in the multiplex network, where each layer is independently affected by a different process. Finally, the last group --- MIM2, is a superposition of IM2 and MIM, i.e., it tackles the problem of spreading multiple processes in multilayer networks that affect actors, not nodes.

The problem in the assessment of state-of-the-art methods lays in the aforementioned diversity, in which the problem of influence maximisation in multilayer networks is defined. Till now, the experimental research on seed selection for influence spreading in multilayer networks has focused mainly on the Independent Cascade Model (ICM).

In~\cite{erlandsson2018seed} authors use Degree Centrality, K-shell decomposition~\cite{kitsak2010superspreaders}, VoteRank~\cite{zhang2016identifying} and \textit{ARL}~\cite{erlandsson2016finding} on multilayer networks build on Facebook and Twitter interactions and conclude that we can achieve the best results in terms of influence maximisation using Degree Centrality. Another recent study~\cite{chen2020maximizing} evaluates Greedy~\cite{kempe2003maximizing}, out-degree, random, as well as \textit{RRE}, \textit{Fair1} and \textit{Fair2} (proposed in the paper) on four complex networks in the context of Fair Seed Allocation problem.

Another work~\cite{wang2022mfearim} introduces a multifactorial evolutionary algorithm \textit{MFEA-RIM\textsubscript{m}}, leveraging multitasking optimisation theory to address the Robust Influence Maximisation problem in multilayer networks. The same team in another work~\cite{wang2022marimmulti} propose a memetic algorithm named \textit{MA-RIM\textsubscript{Multi}} to solve the same problem. In their study, they also evaluate (and adapt to the multilayer network case) approaches like \textit{RSPN-3}~\cite{qiang2018rspn3}, \textit{LAPSO-IM}~\cite{singh2019lapso}, \textit{EA-IM}~\cite{Iacca2021imea}, \textit{CIMMIC}~\cite{XIE2021CIMMIC}, or \textit{DGA}~\cite{JABARILOTF2022GDA} --- however, their research concerns only ICM.

On the other hand, \cite{lu2020nsgaii} describes an application of the Nondominated Sorting Genetic Algorithm II to the problem of Influence Maximization for multilayer networks. The authors of that work compare their method with \textit{LDAG}~\cite{cheng2010ScalableIM}, Degree Discount~\cite{Chen2009DegreeDiscount}, Degree Centrality, Greedy~\cite{kempe2003maximizing}, and PageRank~\cite{page1999pagerank}. Interestingly, they use both ICM and Linear Threshold Model (LTM) in their evaluation and utilise the actor-based approach in the spreading model mechanism.

Another influx of works concerning Influence Maximisation in multilayer networks with ICM proposes methods like Clique Influence Maximisation~\cite{Venkatakrishna2022CIM} and K++ Shell~\cite{Venkatakrishna2023KppShell}. Among methods used in the comparison of these approaches are: Shapley Value-based Influential Nodes~\cite{Narayanam2011spins}, Knapsack Seeding of Networks~\cite{kuhnle2018knapsackseeding}, Deep Learning-Based Influence Maximisation~\cite{li2019disco}, Degree Discount~\cite{Chen2009DegreeDiscount}, K-Shell Decomposition~\cite{shai2007kshell}, and Community-based K-Shell Decomposition~\cite{SUN20211commkshell}. The same research team introduced the Community-Based Influence Maximisation method~\cite{Venkatakrishna2022CBIM}, which was evaluated on both LTM and ICM and compared the majority of formerly mentioned methods in this paragraph plus approaches like Influence Maximisation on Community Structure~\cite{Chen2020imcs} and \textit{IM-ELPR}~\cite{Kumar2021imcshindex}. Nonetheless, the aforementioned team considers the influence maximisation in terms of nodes, which makes their research closer to the MIM or MIM2 groups than IM2, which we considered in this paper.

Subsequent works concerning LTM and multilayer networks are focused on understanding the model behaviour using just one measure for seed selection~\cite{michalski2013convince}, arbitrarily selecting seeds~\cite{zhong2022mltm}, or not selecting seeds at all~\cite{ziolo2022modeling}.

\subsection{Summary and Justification for Tackling the Problem}

With regard to the literature review described above, we decided to focus on the problem of seed selection for the LTM in the problem of budget-constrained influence maximisation on multilayer networks. We also limited the evaluated methods to rank-refining heuristics. That allowed us to focus on a particular group and unify the analysis. Moreover, the fact they do not exploit the properties of the spreading model makes them more versatile. They can also be used in works that, instead of LTM, consider its counterparts, like ICM. With that property, they can be treated as a baseline for more sophisticated seed selection methods. Thus, we adjusted the most popular seed selection strategies defined for single-layer networks like the Degree Centrality, PageRank, K-shell Decomposition, etc. (see Sec.~\ref{sec:seedselection} for details) as well as the novel ones belonging to the aforementioned group (e.g. Community-Based Influence Maximisation) and evaluated them on twelve multilayer networks varying in size, number of layers and topology (see Tab.~\ref{tab:networks_eda}).

The paper is organised as follows: in Sec.~\ref{sec:materials}, we introduce measures, methods, and models utilised in the research. We also formulate the problem that is being tackled in this paper. Next, in Sec.~\ref{sec:seedselection}, we describe seed selection methods employed in the study and how we adapted them to multilayer networks. Sec.~\ref{sec:setup} describes the experimental setup, including evaluated parameters and the list of networks we have used for the evaluation. Finally, we describe the results in Sec.~\ref{sec:results} and Sec.~\ref{sec:final_results}, describing our step-by-step assessment and findings. The paper ends with the conclusions in Sec.~\ref{sec:conclusion}.

\section{Problem Definition}\label{sec:materials}

In this section, we define the problem in a formal way, as well as present all structures, definitions and methods that were necessary for the design and execution of the experimental pipeline.

\subsection{Multilayer Networks}

As it was mentioned in Sec.~\ref{sec:intro}, simple one-layer networks might not be enough to understand and model the spread of influence in the real world. Thus, in this study, we have decided to use multilayer networks~\cite{dickison2016multilayer, kivela2014multilayer}, which allows us to model different interactions in various social cycles. Below, we formally define the network structures used in the following sections.

\begin{theorem}[Network]\label{def:net}
    One can define network as $N=(V,E)$, where $V$ is a non-empty set of nodes and $E$ is a set of edges $(v_1, v_2): v_1, v_2 \in V$.
\end{theorem}

\begin{theorem}[Multilayer Network]\label{def:multilayer_net}
    A tuple $M = (A,L,V,E)$ is called a multilayer network if:
    \begin{itemize}
        \item $A$ is a non-empty set of actors $\{a_1,..., a_n\}$,
        \item $L$ is a non-empty set of layers $\{l_1,..., l_m\}$, 
        \item $V$ is a non-empty set of nodes, $V \subseteq A \times L$,
        \item $E$ is a non-empty set of edges $(v_1^{l}, v_2^{l}): v_1^{l}, v_2^{l} \in V$, and ($v_1^{l_1}, v_2^{l_2}) \Rightarrow E$ then $l_1=l_2$.
    \end{itemize}
\end{theorem}

\begin{theorem}[Multiplex Network]\label{def:multiplex_net}
    Consider a network $M$, which meets Def.~\ref{def:multilayer_net}. If $\forall l \in L: V_{l} \equiv A$, then this kind of network is called multiplex.
\end{theorem}

In other words, a multilayer network can be viewed as a coupled set of networks following Def.~\ref{def:net} so that each actor is represented at most in every layer by the corresponding node ($V \subseteq A \times L$). Multiplex networks only sharpen this property by requiring each actor to be represented in all layers. Moreover, in this work, we do not consider interrelational edges, and links can be produced only within each layer: $(v_1^{l_1}, v_2^{l_2}) \in E \Rightarrow l_1=l_2$. For the sake of simplicity, when referring to particular layers, the aforementioned sets will be marked with a subscript. Thus, $V_l$ and $E_l$ refer to the nodes and edges of layer $l$, respectively.

\begin{figure}[ht]
    \centering
    \includegraphics[width=.50\linewidth]{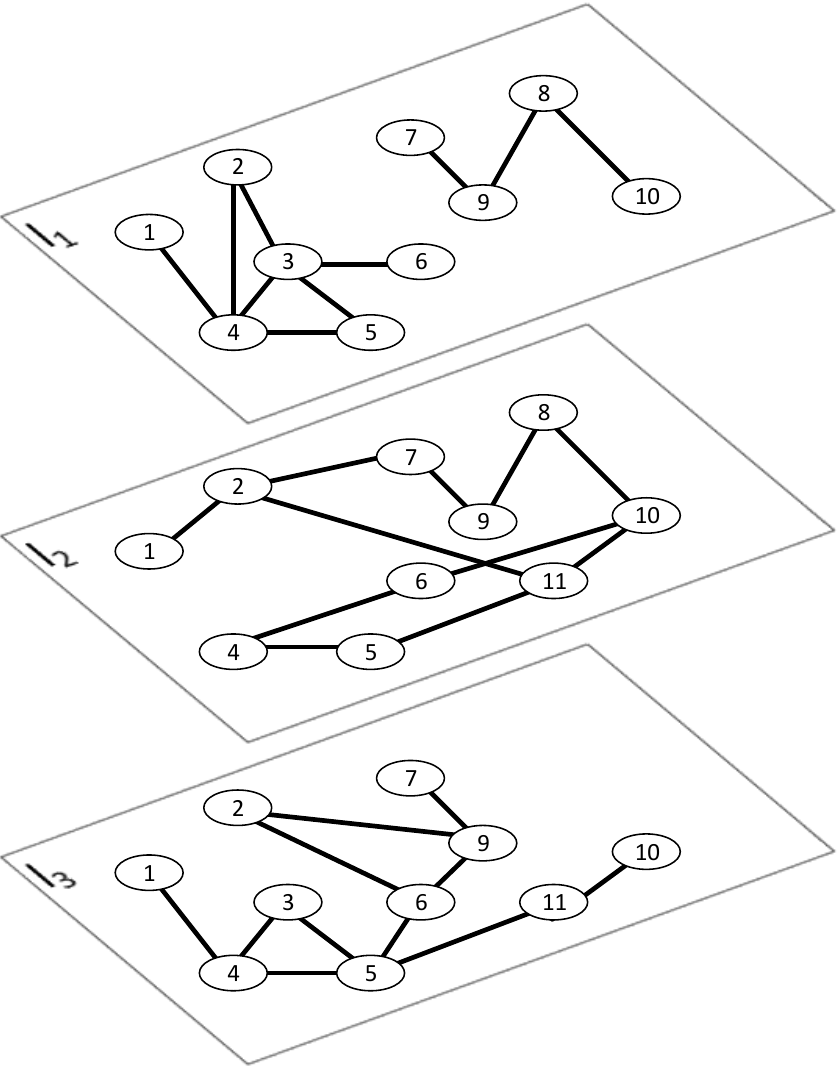}
    \caption{Example of a multilayer network.}
    \label{fig:toy_net}
\end{figure}

To provide a visual example of a multilayer network, we present in Fig.~\ref{fig:toy_net} a toy representative of such a graph. One can note that it contains: 
\begin{itemize}
    \item eleven actors ---  $\{ a_1, a_2, a_3, a_4, a_5, a_6, a_7, a_8, a_9, a_{10}, a_{11} \}$, 
    \item three layers ---  $\{ l_1,l_2, l_3 \}$,
    \item thirty nodes ---  $ \{ v_{1}^{1}, v_{2}^{1}, v_{3}^{1}, ..., v_{1}^{2}, v_{2}^{2}, v_{3}^{2}, ..., v_{1}^{3}, v_{2}^{3}, v_{3}^{3} \}$, 
    \item thirty-two edges --- $\{(v_{1}^{1}, v_{4}^{1}), (v_{2}^{1}, v_{3}^{1}), ..., (v_{1}^{2}, v_{2}^{2}), (v_{2}^{2}, v_{7}^{2}), ..., (v_{1}^{3}, v_{4}^{3}), (v_{2}^{3}, v_{6}^{3})\}$.
\end{itemize}

\subsection{Linear Threshold Model}

Although the LTM~\cite{granovetter1978threshold} has been comprehensively described in the literature, with its variations and generalisations, we will present its most classical form and its extension to multilayer networks. Nonetheless, before introducing the models, the auxiliary constructs must be presented:

\begin{theorem}[Binary Discrete System]\label{def:two_state_sys}
    Let $a$ be an agent\footnote{In our case, an agent can be an actor or node depending on the model.} present in the network structure. At time-step $t$, $a$ can adopt one of two discrete states $x_{a}(t) \in {0, 1}$: \textit{inactive} ($x=0$), and \textit{active} ($x=1$) either through interactions with its neighbours or by preserving $x_{a}(t-1)$. Once $x_{a}(t) = 1$, $a$ remains active eternally, i.e. there is only one state transition allowed: \textit{inactive} $\rightarrow$ \textit{active}.
\end{theorem}

\begin{theorem}[Seeds in the Binary Discrete System]\label{def:seeds}
    Let $S_{t}$ be a set of active agents in a time step $t$ from the system as defined in Def.~\ref{def:two_state_sys}. We denote $S_{0}$ as a seed set, i.e. a set of initially active agents before any dynamics occur in the system. When $S_{t-1} = S_{t}$ the system reaches a steady state.
\end{theorem}

\subsubsection{Linear Threshold Model}\label{subsubsec:ltm}

Following works~\cite{granovetter1978threshold, kempe2003maximizing}, the classical LTM can be formulated as follows:

\begin{theorem}[Linear Threshold Model]\label{def:ltm}
    Consider a system according to Def.~\ref{def:two_state_sys} and Def.~\ref{def:seeds} within a network following Def.~\ref{def:net}. Each node $v$ can be influenced by only its neighbours $U$ given connection weights $w$, so that $\sum_{u \in U} w_{u, v} \leq 1$. Each node $v$ chooses a threshold $\mu_{v}$ randomly from the uniform distribution $\mathcal{U}(0, 1)$, which represents a weighted fraction of neighbours of $v$ that must be active in order to activate $v$. The diffusion process unfolds in discrete steps as follows: a node $v$ at a time step $t$ become active if at $t-1$ influence of its neighbours $U$ exceeded, $\mu_{v}$, i.e. $\sum_{u \in U \cap S_{t-1}} w_{u, v} \geq \mu_{v}$. 
\end{theorem}

\subsubsection{Multilayer Linear Threshold Model}

The question whether LTM in the form as introduced in Def.~\ref{def:ltm} can be directly applied to multilayer networks is ambiguous. In the literature, there are two approaches to solving this issue.

The first one proposes to treat each layer separately so that a given actor can be active on some layers and at the same time, be not active on other ones. For instance, works like~\cite{Venkatakrishna2022CIM, Venkatakrishna2022CBIM} adopt this attitude in the problem of influence maximisation. According to~Sec.\ref{subsec:infmaxmln}, this framework can be treated as a variation of MIM. However, we believe that the applicability of that approach is relatively low. Suppose an actor is activated at one layer and still is not at another. In that case, one can make two independent experiments treating each network layer independently --- there is no reason to use a multilayer approach. Moreover, in that case, the nodes, not the actors, are the subject of diffusion, so the seed set consists of nodes picked from each layer independently. This approach is nonetheless applicable to problems such as simulating multiple marketing campaigns at the same time. However, if we would like to examine a single spreading process in the multilayer system, it is less useful since decisions made on one layer do not influence decisions on other layers.

The second approach assumes that the actors, not nodes, are the subject of the diffusion process. Therefore, a natural consequence of this condition is to define what it means that an actor is (or is not) active, how it relates to layers (where respective nodes represent it), and where spreading dynamics take place. One of the possible answers to these questions is given in~\cite{zhong2022mltm} by introducing so-called protocols. These are aggregation functions that determine whether to activate an actor by signals from layers where it is represented. In this work, we adopted that approach. Nonetheless, since this study is intended to be an experimental one, we refined the model introduced in~\cite{zhong2022mltm} as follows:
\begin{itemize}
    \item the heterogeneity (in the sense that each node picks randomly threshold $\mu_{v} \sim \mathcal{U}(0, 1)$) has been neglected; instead, we impose one value of $\mu$ for all actors,
    \item consequently, heterogeneity of the protocol functions properly for the actors has been transformed into homogeneity over the entire network,
    \item a requirement for the multiplicity (Def.~\ref{def:multiplex_net}) of the network has been relaxed; thus, any multilayer network can be a medium for the influence propagation model.
\end{itemize}
The justification for our amendments was an intention to simplify the experimental setup and limit the number of variables, thereby increasing the observability of the impact of $\mu$ and protocols on the model performance. Moreover, since this work is purely experimental, lifting the restriction that the network must be a multiplex allows us to test the model on virtually any graph.

\begin{theorem}[Multilayer Linear Threshold Model]\label{def:mltm}
    Consider an actor $a$ in the multilayer network $M = (A, L, V, E)$ represented in $K$ layers (i.e. $K \subseteq L$) by a set of its representative nodes: $V_{a} = \{..., v_{a}^{k}, ...\} : k \in K$ and a system according to Def.~\ref{def:two_state_sys}, \ref{def:seeds}. Each node $v$ can be influenced only by its neighbours $U$ given connection weights $w$, so that $\forall u \in U: \space {w_{u, v} = degree(v)^{-1}}$ and $\sum_{u \in U} w_{u, v} = 1$. The diffusion process unfolds in discrete steps as follows: node $v$ at time step $t$ receives input $y$ according to threshold $\mu$:
    \begin{equation*}
        y_{v}(t) =
        \begin{cases}
          1,  & \text{if \space} \sum_{u \in U \cap S_{t-1}} w_{u, v} > \mu \\
          0,  & \text{otherwise}
        \end{cases} 
    \end{equation*}
    Let $ y_{a}(t) = |K|^{-1} \sum_{k \in K} y_{v}^{k}(t)$ be the mean input of the actor $a$ in the time step $t$. Then, given $y_{a}(t)$, one can determine the state of the actor with a protocol function (Def.~\ref{def:proto}).
\end{theorem}

\begin{theorem}[Protocol Function]\label{def:proto}
    According to~\cite{zhong2022mltm}, state of the actor $a$ of the multilayer network $M = (A, L, V, E)$ in the time step $t$ is determined by a function: 
    \begin{equation*}
        x_{a}(t) =
        \begin{cases}
          1,  & \text{if \space} y_{a}(t) \geq \delta \text{\space or \space} x_{a}(t-1) = 1 \\
          0,  & \text{otherwise}
        \end{cases} 
    \end{equation*}
    Where $\delta \in [\frac{1}{L}, 1]$ is a parameter.
\end{theorem}

From the practical point of view, we can notice that, if compared to the classic LTM, the model introduced above (MLTM) adds another threshold by a notion of the protocol. That function behaves like a quantiser that determines whether the actor obtained "sufficient influence" for activation by evaluating the influence obtained by its representative nodes. If so, it will be activated on all layers. Conversely, none of the nodes will be set to active if the mean input is lower than $\delta$. Thus, the activation conditions can be controlled both with $\mu$ and $\delta$. It is harder to activate actors for higher values of $\delta$ and $\mu$, than if they are lower.

\begin{figure}[ht]
    \centering
	\includegraphics[width=1\linewidth]{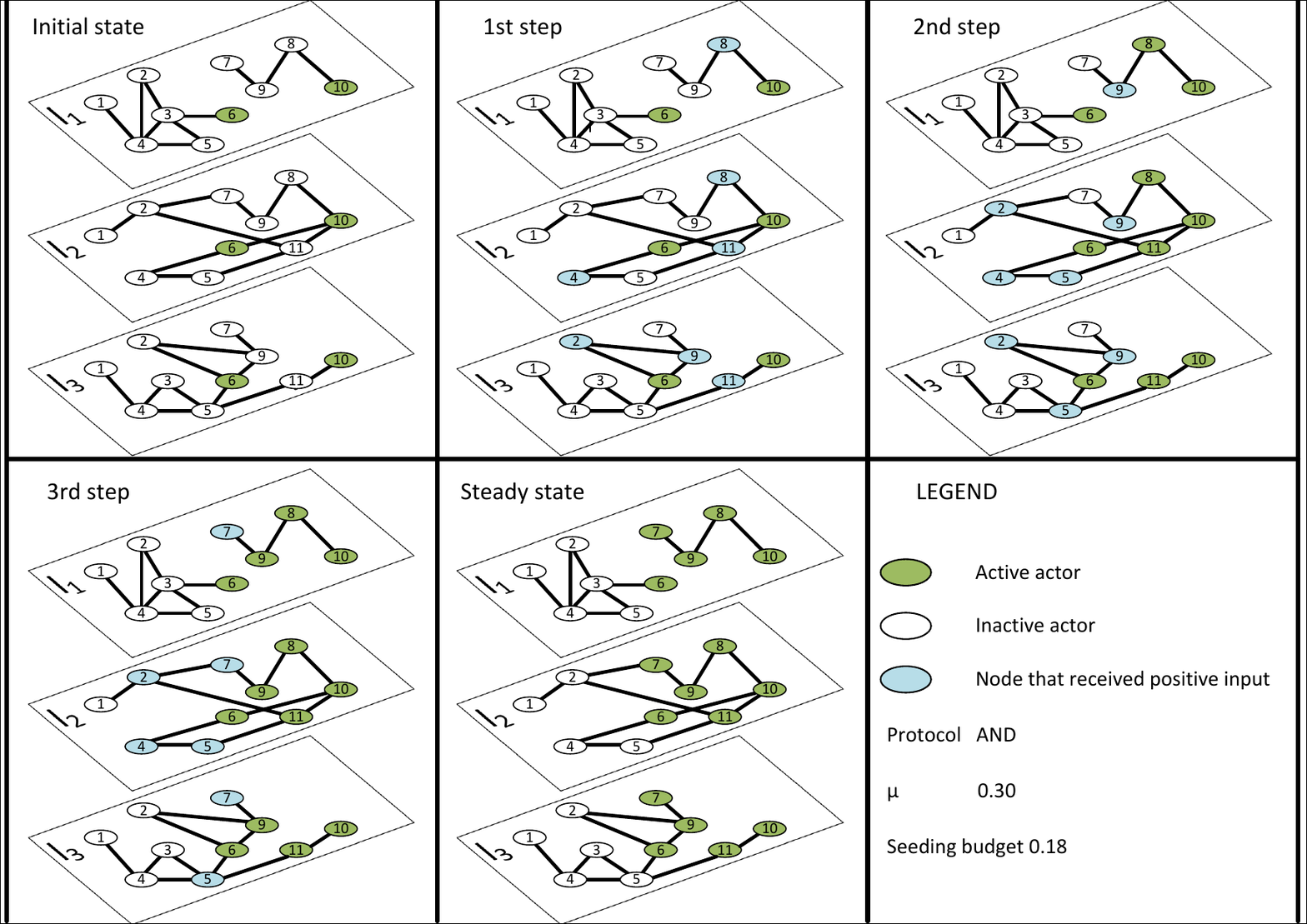}
	\caption{Example of spreading of MLTM through toy network with protocol $AND$.}
    \label{fig:ltm_example_and}
\end{figure}

\begin{figure}[ht]
	\centering
	\includegraphics[width=1\linewidth]{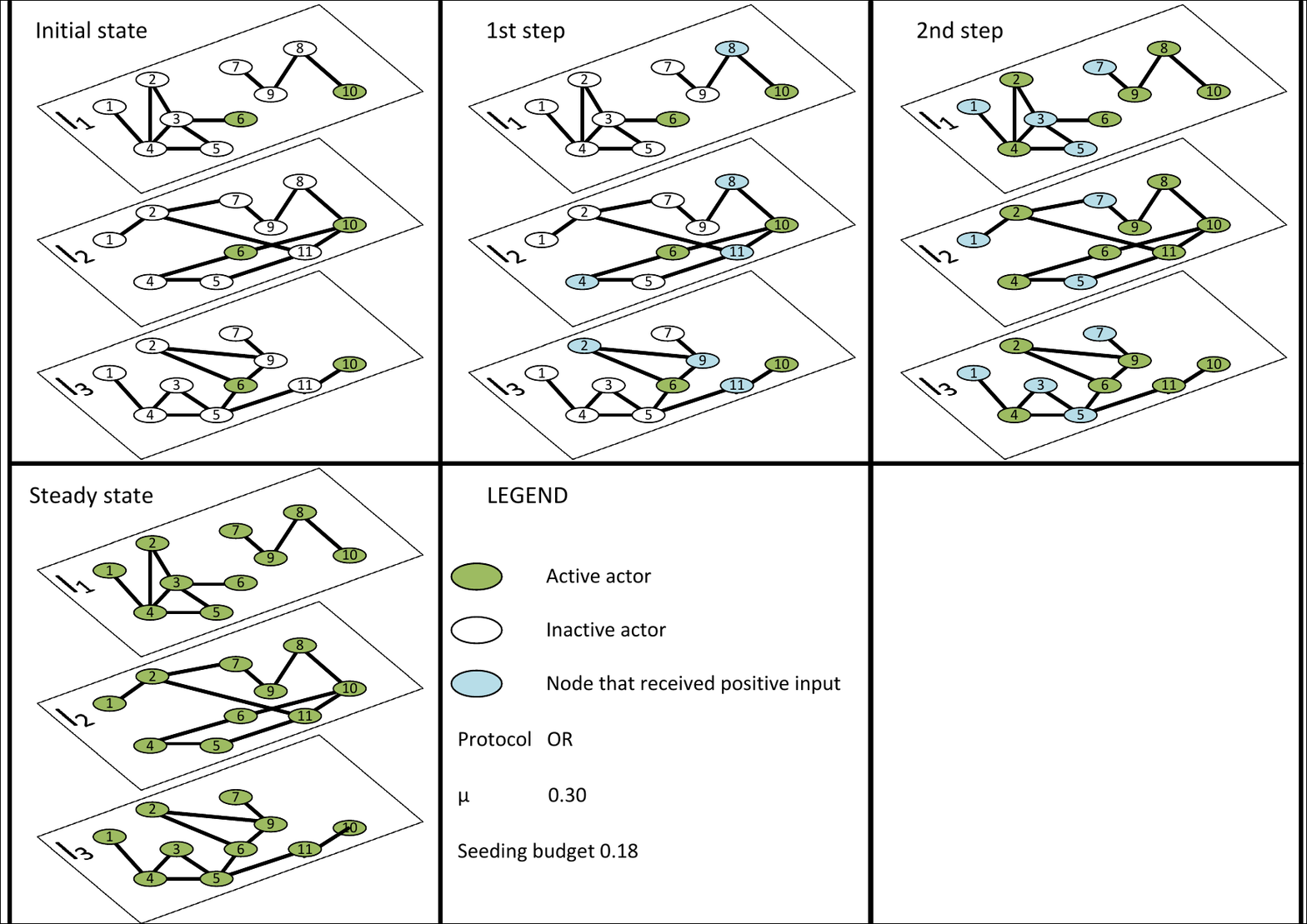}
	\caption{Example of spreading of MLTM through toy network with protocol $OR$.}
    \label{fig:ltm_example_or}
\end{figure}

Regarding the above, we decided to examine two extreme cases: $\delta = 1$ (hereinafter called $AND$) and $\delta = \frac{1}{L}$ (hereinafter called $OR$). In the former case, an actor gets activated if it receives sufficient influence on all layers where it is represented, and conversely, in the latter, where sufficient input in at least one layer is enough for activation. In Fig.~\ref{fig:ltm_example_and} and Fig.~\ref{fig:ltm_example_or}, we present differences in these protocols in the example of diffusion under MLTM in the toy network.

\subsubsection{Influence under MLTM}

The issue of measuring the influence under a particular spreading model is thoroughly covered in the literature, as shown in Sec.~\ref{subsec:spreading_performance}. The metrics employed and the properties of the objective function enable a theoretical assessment of the effectiveness of methods designed to solve the influence maximisation problem. Following the seminal work of~\cite{kempe2003maximizing}, the influence function $\sigma: S \rightarrow \mathbb{N}$ is usually formulated as a function that maps a set of agents into a natural number. This represents the expected number of active agents at the steady state when $S$ is used as the seed set. For the classic LTM, it is proven that $\sigma$ is submodular, which provides an approximation guarantee of heuristics such as greedy hill climbing~\cite{kempe2003maximizing}. However, the authors of MLTM do not address this aspect in their work~\cite{zhong2022mltm}. In the following section, we aim to resolve this gap.

A function is said to be submodular if it satisfies the diminishing marginal returns property, meaning that the incremental benefit of adding a new element decreases as the set grows. In the context of the considered problem, this implies that the marginal benefit of adding a new actor to an existing seed set diminishes as the seed set becomes larger.

\begin{theorem}[Submodular Function]\label{def:submodularity}
A function $f$ is submodular if for any sets $\mathcal{S}\sp{\prime} \subseteq \mathcal{S}$ and any element $ a \notin \mathcal{S} $, the following inequality holds:
    \begin{equation*}
        f(\mathcal{S}\sp{\prime} \cup \{a\}) - f(\mathcal{S}\sp{\prime}) \geq f(\mathcal{S} \cup \{a\}) - f(\mathcal{S})
    \end{equation*}
\end{theorem}

As noted above, the key difference between the classic LTM and its multilayer counterpart lies in the aggregation operation (i.e. a protocol function), where the additional threshold $\delta$ is employed. That, however, makes the influence spread non-submodular. We can prove it in the following example. Consider a diffusion within a network from Fig.~\ref{fig:toy_net} under MLTM with $\mu = 0.5$ and $\delta = 1/L$ (i.e. variant $OR$). Then, for a seed set $S_{0} = \{a_{6}, a_{11}\}$, the value of the influence function $\sigma(S_{0}) = 3$, because $S_{t \rightarrow \infty}=\{a_{6}, a_{11}, a_{10}\}$. From the other hand, the diffusion for its subset $S_{0}' = \{a_{11}\}$, results in $\sigma(S_0')=2$, because $S'_{t \rightarrow \infty}= \{a_{11}, a_{10}\}$. However, if these seed sets get extended by the additional actor $a_5$, the following occurs: $\sigma(S_0 \cup \{a_5\}) = 11$, because $S \cup \{a_5\}_{t \rightarrow \infty } = \{a_{6}, a_{11}, a_{5}, a_{4}, a_{10}, a_{1}, a_{3}, a_{2}, a_{9}, a_{7}, a_{8}\}$ and $\sigma(S'_0 \cup \{a_5\}) = 3$, because $S' \cup \{a_5\}_{t \rightarrow \infty } = \{a_{5}, a_{11}, a_{10}\}$. Finally, we get $\sigma(S\sp{\prime} \cup \{a_{5}\}) - \sigma(S\sp{\prime}) = 3 - 2 = 1$ and $ \sigma(S \cup \{a_{5}\}) - \sigma(S) = 11 - 3 = 8$ which results in $1 < 8$. Thus, the inequality from Def.~\ref{def:submodularity} is not met.


\subsection{Influence Maximisation Problem}\label{subsec:infmax}

Having proven that the classically formulated influence function under MLTM is not submodular, we are not constrained to use it as a measure of diffusion effectiveness. Therefore, we present below how this paper treads the influence maximisation problem and what metrics are employed to assess the methods used in this study.

\begin{theorem}[Budget Constrained Influence Maximisation under MLTM]\label{def:infmax}
    Let $\mathcal{S}$ be a family of sets of cardinality $s$ over actors of the multilayer network $M$ in the sense that $\mathcal{S} \subseteq powerset(N)$. Let $\sigma: \mathcal{S} \rightarrow \mathbb{R^{+}}$ be an arbitrary function that maps a set of actors to a number denoting the efficiency of using it as a seed set to trigger the diffusion under MLTM. An influence maximisation for seeding budget of size $s$ is a problem of finding $ S_{0}: \arg\max(\sigma) = S_{0} \wedge |S_{0}| \leq s$.
\end{theorem}

Following that, we introduce a metric of influence used in this work --- Gain ($G$). Another auxiliary measure (which, however, does not meet Def.~\ref{def:infmax}) is Diffusion Length ($DL$). It can be used to decide which seed selection methods are more efficient if they exhibit similar $G$. Tab.~\ref{tab:indicators} presents their formulas and descriptions.

\begin{table}[ht]
    \caption{Indicators of diffusion efficiency taken into account.}
    \begin{tabular}{l|l|p{8.7cm}}
    Symbol & Indicator  & Description \\ \hline \hline
    $DL$ & Diffusion Length & Number of simulation steps passed until a steady state is reached. \\ \hline
    $G$ &  Gain & Influence under MLTM achieved by a seed set $S_0$ denoted as a proportion of the actors activated during the diffusion and the total number of activatable actors: $G = 100 \cdot \frac{|S_{DL} - S_{0}|}{|A - S_{0}|}$, $G \in [0, 100]$. \\
    \end{tabular}
    \label{tab:indicators}
\end{table}

Nonetheless, it is worth noting that if $G$ is intuitively expected to be high, the value of $DL$ may be open to discussion. In some cases, a great diffusion length is desired. For instance, in viral marketing, an advertiser might prefer to have the message circulating within society for an extended period of time. On the other hand, systems that rely on efficient message passing, such as computer networks, aim to transmit information in the shortest possible time. In this paper, we assume the latter case. Thus, despite the secondary role of $DL$ in evaluating seed selection methods, we consider its lower values to be preferable to higher ones.

\section{Seed Selection Methods Used in the Study}\label{sec:seedselection}

During the study, we evaluated the following heuristics that can be classified into the rank refined group: Community-Based Influence Maximisation~\cite{Venkatakrishna2022CBIM}, Clique Influence Maximisation~\cite{Venkatakrishna2022CIM}, Degree Centrality, Degree Discount~\cite{Chen2009DegreeDiscount}, K-shell Decomposition~\cite{shai2007kshell}, K++ Shell~\cite{Venkatakrishna2023KppShell}, Neighbourhood Size~\cite{magnani2011ml}, Neighbourhood Size Discount, PageRank~\cite{page1999pagerank}, and VoteRank~\cite{zhang2016identifying}. We also adopted intuition presented in a work of~\cite{kempe2003maximizing} and included in the evaluation greedy hill climbing and random choice as the potential upper and lower boundaries for the other methods (as mentioned above since MLTM is not submodular, we cannot claim that the greedy hill climbing gives us real upper bound approximation~\cite{kempe2003maximizing}). Each seed selection method was designed to return a list of all actors from the network, sorted descending by their position in the ranking. Keeping this in mind, one can note that a budget of seeds is a number of the first $s$ actors from the ranking list.

Moreover, due to the specific nature of the problem considered in this study, we had to extend some methods to the case of multilayer networks (i.e. K-shell Decomposition, PageRank, and VoteRank). Thus, we implemented them in two ways. The first one assumed that initial computations are done at the node level and then aggregated for each actor. The second approach took the perspective of the actor from the very beginning. Nonetheless, despite being developed for multilayer networks, some heuristics (e.g. Community-Based Influence Maximisation) addressed the Multiple Influence Maximisation problem, which is not considered in this study. Hence, we also adapted them to the problem described in Sec.~\ref{sec:materials}. For the sake of simplicity, we based experiments on unweighted, undirected networks without self-loops. Therefore, all seed selection methods were defined to work with such graphs (e.g. for Degree Centrality, we did not differentiate out-degree from in-degree).

\subsection{Community-Based Influence Maximisation (\textit{cbim})}

The Community-Based Influence Maximisation proposed by~\cite{Venkatakrishna2022CBIM} consists of two phases. First, small communities are identified using the function based on Dice Neighbourhood Similarity~\cite{dice1945similarity}. Then, they are consolidated to form larger ones to enhance the community division. In the second phase, an Edge Weight Sum (proposed by the authors of \textit{cbim}) is computed for each node within a community. Finally, nodes are ranked based on these values, and the seed set is selected from the communities using a quota-based approach.

Despite being addressed to multilayer networks, this method finds the most influential nodes, not actors, which does not address the problem considered in this work. Therefore, our amendment was to transform rankings computed for each layer separately by computing a weighted average position (by the layer size) of actors in the rankings and then yield a final rank of the actors by those values. Instead of Edge Weight Sum, we also used a Katz Centrality~\cite{katz1953centrality}, which is a similar algorithm in order to speed up the computations.

\subsection{Clique Influence Maximisation (\textit{cim})}

This method is somehow similar to \textit{cbim}. It was also designed for multilayer networks, and similarly, its objective is to find a seed set of nodes, not actors. The main difference is that instead of dividing the network into communities, it seeks maximal cliques. Then it ranks nodes by their degree and picks iteratively the highest ranked nodes from each clique starting from the largest~\cite{Venkatakrishna2022CIM}. For \textit{cim}, our amendment was also to convert rankings computed for each layer to those for actors based on their weighted average position.

\subsection{Degree Centrality (\textit{deg-c})} 

The next metric we used was the Degree Centrality. It was computed as follows: for each actor $a$, count nodes connected to the actor $a$ on each layer~\cite{magnani2011ml}. Degree Centrality is the number of those nodes.

\subsection{Degree Centrality Discount (\textit{deg-c-d})}\label{subsec:deg-c-d}

Degree Centrality Discount approach was initially defined in~\cite{Chen2009DegreeDiscount} as a simple and effective heuristic for ICM. It was based on the concept that highly connected nodes may already have a high level of influence due to their extensive connections, and thus, if they are picked to the seed set, the contribution of their neighbours should be discounted so that seeds are equally distributed through the graph. We have modified it first by neglecting the impact of the spreading model on the discounting process, so that it is based only on the structural properties of the graph. The second modification was using Degree Centrality in the form defined for multilayer networks~\cite{magnani2011ml}, which allowed the method to yield a seed set of the actors.

\subsection{Greedy Method (\textit{greedy})}\label{subsec:greedy}

We used a straightforward greedy hill climbing routine, similarly to~\cite{kempe2003maximizing}. This approach assumes the iterative building of a seed set that results in the best $G$ increment until the budget is spent. We start by evaluating the model's performance with a one-element seed set. We choose an actor $a_{i}$ that when used, results in the best $G$. Then, we seek another actor $a_{j}$ that, used with $a_{i}$ as seeds, results in the best $G$ or at least the same $G$ but shorter $DL$. If a few actors result in the same increment of $G$ and the reduction of $DL$, we choose a random one from them. We repeat this step with the rest of the actors, and as a result, we obtain a greedily chosen seed set.

\subsection{K-shell Decomposition (\textit{k-sh} and \textit{k-sh-m})}

Nodewise extension (\textit{k-sh}) was designed as follows. For each layer, perform K-shell Decomposition~\cite{shai2007kshell} and compute the degree of nodes. Then, sort nodes in descending order based on two conditions: the shell they belong to and their degree (respectively, the deeper and the higher, the better). The final ranking list is computed as the weighted average position of an actor in the rankings of layers where it is represented.

The second way we adapted the K-shell Decomposition to multilayer networks utilised the actorwise approach (\textit{k-sh-m}). Compared to the implementation by~\cite{shai2007kshell}, we used the actor's neighbourhood size instead of the node's degree. That follows the intuition of the original algorithm --- for a single-layer network, the neighbourhood size of an actor is similar to the degree of its corresponding node. That approach allows graphs to be decomposed into shells that encompass all layers and contain actors, not nodes. Finally, we compute the ranking by sorting actors by the shell to which they were assigned and their degrees.

\subsection{K++ Shell (\textit{kpp-sh})}

The next method we used was the K++ Shell introduced by~\cite{Venkatakrishna2023KppShell} to identify a set of seeds within the multilayer networks. This algorithm, inspired by K-shell, iteratively prunes nodes from the network based on their degree and adds it to shells, but, in addition, introduces reward points for neighbours of nodes pruned in the previous step of the routine. Considering the node's shell number and reward points, the K++ Shell algorithm addresses the limitation of exclusively choosing nodes from the highest bucket. It also utilises a community detection approach to create the ranking list by picking nodes iteratively from each community according to the seeding budget. However, it operates on nodes similarly to \textit{cbim} and \textit{cim}. We adapted it to the problem considered in the paper similarly by taking the weighted average position (by the layer size) of actors in the rankings for each layer.

\subsection{Neighbourhod Size (\textit{nghb-1s} and \textit{nghb-2s})}

The Neighbourhood Size of an actor is the number of actors it is connected to, regardless of the connection multiplicity, resulting from layers where interactions between neighbours occur~\cite{magnani2011ml}. In experiments, we used two versions of this metric. The first was based on a one-hop neighbourhood (hereinafter \textit{nghb-1s}) and based on a two-hop setting (\textit{nghb-2s}). It is worth noting the difference between this metric and Degree Centrality: neighbourhood size does not account for the number of links connecting a pair of actors (which may be repeated due to the network’s multilayer structure), whereas this is relevant in the latter method.

\subsection{Neighbourhood Size Discount (\textit{nghb-sd})}

We also decided, based on the Degree Centrality Discount~\cite{Chen2009DegreeDiscount}, to propose a Neighbourhood Size Discount by utilising Neighbourhood Size. The method was similar to one described in Sec.~\ref{subsec:deg-c-d}, but instead of the Degree Centrality, we used the one-hop Neighbourhood Size of the actor.

\subsection{PageRank (\textit{p-rnk} and \textit{p-rnk-m})}

The node-wise ranking (\textit{p-rnk}) was obtained by computing PageRank, as introduced by its authors in~\cite{page1999pagerank}, on each network's layer. Then, the final ranking list was created by ordering actors by their mean position in layers rankings.

The second implementation (\textit{p-rnk-m}) adapted the original PageRank algorithm to the case considered in this work differently. Namely, we squeezed a multilayer network into a single-layer graph to preserve all actors. Then, we created edges between actors if at least one layer of the original network had such a link. Then, an output of a PageRank algorithm computed on a squeezed network was a final ranking list.

\subsection{Random Choice (\textit{random})}

We also examined the random choice of actors. A function \lstinline{random.random()} from a standard Python library was used to generate a seed set. In order to observe the impact of this selection strategy on diffusion efficiency, the experiment was repeated 20 times for each evaluated combination of parameters.

\subsection{VoteRank (\textit{v-rnk} and \textit{v-rnk-m})}

Our first approach (subsequently \textit{v-rnk}) to adapt VoteRank~\cite{zhang2016identifying} was implemented similarly as in \textit{k-sh}. That is, for each layer, the VoteRank algorithm was executed, resulting in a ranking of nodes that got votes. That list, however, did not contain all nodes, so to satisfy conditions where an extensive seed set was needed, we extended it by nodes that did not get any votes (i.e. were not recognised by the VoteRank as central agents in the network). The final ranking of actors was obtained by computing the weighted mean position of actors in layer-wise ranks.

Another implementation (\textit{v-rnk-m}) was based on a deeply modified VoteRank algorithm so that it could be applied directly to the multilayer networks, i.e. actors, not nodes, are the subject of the algorithm. Therefore, instead of Degree Centrality, we used the Neighbourhood Size as a denominator in the voting potential formula. We also considered connections between actors regardless of layers where that relation takes place and their multiplicity. Similarly to the original algorithm, our implementation returns the ranking of actors who got at least one vote. Hence, to obtain a complete list, we extended it by the actors that did not obtain any ballots.

\subsection{Summary}

\begin{figure}[ht]
    \centering
    \includegraphics[width=.7\linewidth]{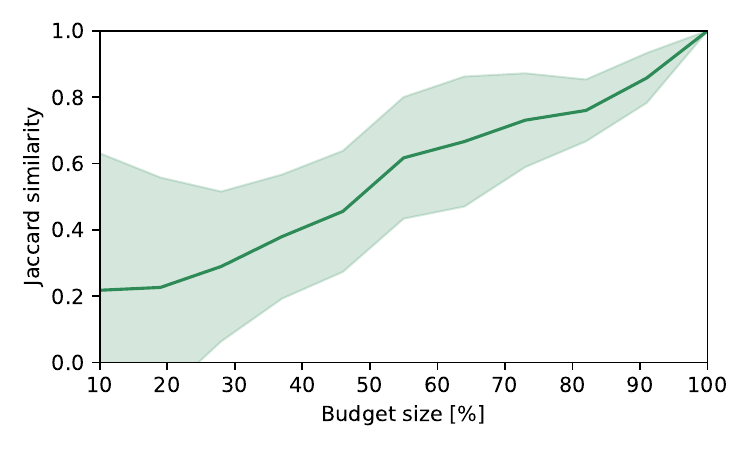}
    \caption{Mean pairwise Jaccard similarity with standard deviation between seed sets constructed with the ranking lists from Tab.~\ref{tab:rankings_toy_net} in a function of the budget size.}
    \label{fig:similarities_rankings_toy_net}
\end{figure}

Having introduced the seed selection methods employed in the study, we show how they work on a toy example. Nonetheless, we refer curious readers to supplementary materials with a source code of their implementation (see Sec.~\ref{subsec:implementation}). Tab.~\ref{tab:rankings_toy_net} presents rankings of actors obtained with the aforementioned seed selection methods (without random choice and greedy method, which depend on the spreading model). Please note differences, especially between heuristics, that have been implemented in two ways. As stated earlier, such rankings are being used to construct seed sets. Fig.~\ref{fig:similarities_rankings_toy_net} shows how similar they are regarding the budget size. As observed, for smaller budgets, their average similarity does not exceed $30\%$, but as $s$ increases, similarity grows. This confirms the significant impact of $s$ on different seed selection methods.

\begin{table}[ht]
    \centering
    \caption{Positions of each actor in a toy network (Fig.~\ref{fig:toy_net}) as ranked by the given seed selection method. Green-coloured cells mark the \first{first}, \second{second}, and \third{third} actor in the ranking, respectively.}
    \begin{tabular}{l||rrrrrrrrrrr}
    \multicolumn{1}{c||}{\multirow{2}{*}{S.S.M.}} & \multicolumn{11}{c}{Actor's ID} \\
    & 1 & 2 & 3 & 4 & 5 & 6 & 7 & 8 & 9 & 10 & 11 \\ \hline \hline
    cbim & 9 & 5 & \second{2} & \first{1} & 7 & 8  & 10 & \third{3} & 4 & 11 & 6 \\ \hline
    cim & 11 & 7 & \third{3} & \second{2} & 8 & 4  & 9  & 5  & 6 & 10 & \first{1} \\ \hline
    deg-c & 11 & \third{3} & 5 & \first{1} & \second{2} & 6 & 9 & 10 & 4 & 7 & 8 \\ \hline
    deg-c-d & 10 & 4 & 6 & \first{1} & \second{2} & 11 & 7  & 8  & \third{3} & 5  & 9 \\ \hline
    k-sh & 11 & \third{3} & \first{1} & \second{2} & 5 & 8 & 10 & 6 & 7 & 9 & 4 \\
    k-sh-m & 11 & \third{3} & 4 & \first{1} & \second{2} & 5 & 9 & 10 & 6 & 7 & 8 \\ \hline
    kpp-sh & 8  & 6 & \third{3} & 7 & \second{2} & 10 & 11 & \first{1}  & 4 & 9  & 5 \\ \hline
    nghb-2s & 7 & \first{1} & 6 & 5 & 8 & \second{2} & 9 & 11 & \third{3} & 10 & 4 \\
    nghb-1s & 9 & \first{1} & 4 & \third{3} & 5 & \second{2} & 10 & 11 & 6 & 7 & 8 \\ \hline
    nghb-sd & 7 & \first{1} & 8 & \third{3} & 4 & \second{2}  & 9  & 10 & 5 & 6  & 11 \\ \hline
    p-rnk & 11 & 6 & \second{2} & \third{3} & 7 & 10 & 9 & 5 & 4 & 8 & \first{1} \\
    p-rnk-m & 11 & \first{1} & 6 & \third{3} & 5 & \second{2} & 10 & 9 & 4 & 7 & 8 \\ \hline
    v-rnk & 8 & 6 & 4 & \first{1} & 7 & 9 & 11 & 5 & \second{2} & 10 & \third{3} \\
    v-rnk-m & 10 & \first{1} & 9 & 5 & 6 & \second{2} & 7 & 8 & 4 & \third{3} & 11 \\
    \end{tabular}
    \label{tab:rankings_toy_net}
\end{table}

\section{Experiment Setup}\label{sec:setup}

In order to provide reliable results, a proper experimental setup was defined. In this chapter, we describe the evaluated set of parameters, the dataset used, the implementation of the simulation environment and, most importantly, the methodology of experiments.

\subsection{Data Used}

The efficiency of a spreading model varies depending on the medium of propagation --- a network. Moreover, as described in Sec.~\ref{sec:intro}, dynamics in multilayer networks are usually not similar to dynamics in single-layer networks. Therefore, selecting proper data to perform experiments is very important. Hence, we decided to use both real and artificially generated graphs. A summary of the networks used in this study, along with a description of the domain context of the data used to construct each network, is provided in Tab.~\ref{tab:networks_eda}. All artificial networks were created with the \textit{multinet} library~\cite{magnani2021analysis}. The real-world networks were read directly from the raw data, except for the \textit{timik} dataset. Namely, due to its big size, we have limited the network to the interactions from the first quarter of 2009. 

\begin{table}[ht]
    \caption{Networks used in experiments with their basic parameters shortlisted.}
    \begin{tabular}{lrrrrp{5.7cm}}
    Name & Layers & Actors & Nodes & Edges & Note \\ \hline \hline
    arxiv & 13 & 14,065  & 26,796  & 59,026 & Coauthorship network obtained from articles published on the "arXiv" repository~\cite{dedomenico2015arxiv}. \\
    aucs & 5 & 61 & 224 & 620 & A graph of interactions between employees of \textbf{A}arhus \textbf{U}niversity, Department of \textbf{C}omputer \textbf{S}cience~\cite{rossi2015aucs}. \\
    ckmp & 3 & 241 & 674 & 1,370 & A network depicting diffusion of innovations among physicians~\cite{coleman1957ckmp}. \\
    eutr-A & 37 & 417 & 2,034 & 3,588 & The European air transportation network~\cite{cardillo2013eutransportation}. \\
    lazega & 3 & 71 & 212 & 1,659 & A network of various types of interactions between staff of a law corporation~\cite{snijders2006lazega}. \\
    timik & 3 & 61,702 & 102,247 & 881,676 & A graph of interactions between users of the virtual world platform for teenagers~\cite{jankowski2024timik}. \\ \hline
    er-2 & 2 & 1,000 & 2,000 & 5,459 & Erd\H{o}s-R\'{e}nyi artificial network~\cite{er-model}. \\
    er-3 & 3 & 1,000 & 3,000 & 7,136 & Erd\H{o}s-R\'{e}nyi artificial network~\cite{er-model}. \\
    er-5 & 5 & 1,000 & 5,000 & 15,109 & Erd\H{o}s-R\'{e}nyi artificial network~\cite{er-model}. \\ \hline
    sf-2 & 2 & 1,000 & 2,000 & 4,223 & Scale-free artificial network~\cite{sf-model}. \\
    sf-3 & 3 & 1,000 & 3,000 & 5,010 & Scale-free artificial network~\cite{sf-model}. \\
    sf-5 & 5 & 1,000 & 5,000 & 10,181 & Scale-free artificial network~\cite{sf-model}. \\
    \end{tabular}
    \label{tab:networks_eda}
\end{table}

The experiments were executed in two stages. During the first one, we wanted to rank seed selection methods preliminarily. Therefore, we utilised smaller networks (i.e. \textit{aucs}, \textit{ckmp}, \textit{eurt-A}, \textit{lazega}, \textit{er-2}, \textit{er-3}, \textit{er-5}, \textit{sf-2}, \textit{sf-3}, \textit{sf-5}) at that phase. Conversely, at the second stage, we were evaluating the five top-so-far heuristics on larger networks (i.e. \textit{arxiv}, \textit{timik}). Ultimately, we used twelve graphs in the study with sizes up to $880,000$ edges, $100,000$ nodes, $60,000$ actors, and $37$ layers.

During the first stage, we have also used single-layer networks to verify the correctness of the extension of some seed selection methods (for instance, \textit{deg-c} and \textit{nghb-s} should yield the same ranking for the single-layer network). However, we decided not to report these results to maintain the briefness of the article. Interested readers can find them in the supplementary materials (see. Sec.~\ref{subsec:implementation}).

\subsection{Evaluated Parameters of the MLTM}

We considered various node activation thresholds ($\mu$), beginning from $0.1$ and finishing at $0.9$, with the step equals $0.1$. It is good to recall that we did not introduce heterogeneity in the model --- one value of $\mu$ was applied to all actors in the network and, consequently, all nodes. That operation allowed us to discard an additional degree of freedom, which could complicate result analysis.

After the initial series of experiments, we observed that the protocol $AND$ is much more demanding than $OR$. That is, values of $s$ sufficient to activate the entire network using protocol $OR$ were not high enough to trigger spreading when the $AND$ strategy was used to aggregate inputs from layers. Therefore, we decided to match different ranges of seeding budgets with each of the protocols (see Tab.~\ref{tab:parameters}). Moreover, we did not see any sense in experiments with large budgets --- that cases are distant from real-world situations. Hence, budget values were chosen to reflect cases where only an actor minority is initially activated.

We also had to determine another factor --- how to spend the seed set. Despite the existence of sophisticated mechanisms like sequential seeding (Sec.~\ref{subsubsec:seed_activation_strategy}), we did not use it due to initial assumptions about the fundamental character of this study. Consequently, all actors selected as seeds were activated at the beginning of the simulation.

Finally, each seed selection method had $252$ combinations of the MLTM parameters to evaluate on the base networks, plus the top five of them were run on bigger networks. That resulted in $41,328$ designed experiments. All considered parameter ranges are presented in Tab.~\ref{tab:parameters}.

\begin{table}[ht]
    \centering
    \caption{Values of parameters evaluated in the study.}
    \begin{tabular}{p{2.5cm}||cc}
    Parameter & \multicolumn{2}{c}{Evaluated values} \\ \hline \hline
    Protocol ($\delta$) & $OR$ & $AND$ \\
    Seed set size ($s$) [\%] & \begin{tabular}{cc}$\{1, 2, 3, 4, 5, 6, 7, $\\ $8, 9, 10, 15, 20, 25, 30\}$\end{tabular} & \begin{tabular}{cc}$\{15, 20, 25, 30, 31, 32, 33,$\\ $ 34, 35, 36, 37, 38, 39, 40\}$\end{tabular} \\
    Treshold ($\mu$) & \multicolumn{2}{c}{$\{0.1, 0.2, 0.3, 0.4, 0.5, 0.6, 0.7, 0.8, 0.9\}$} \\
    Phase 1 networks & \multicolumn{2}{c}{\{aucs, ckmp, eutr-A, lazega, er-2, er-3, er-5, sf-2, sf-3, sf-5\}} \\
    Phase 2 networks & \multicolumn{2}{c}{\{arxiv, timik\}} \\
    S.S.M. & \multicolumn{2}{p{8.5cm}}{\centering \{cbim, cim, deg-c, deg-c-d, greedy\tablefootnote{Due to high computational complexity of greedy method we had to limit evaluation only to real networks.}, k-sh, k-sh-m, kpp-sh, nghb-2s, nghb-1s, nghb-sd, p-rnk, p-rnk-m, random\tablefootnote{Due to nondeterministic character of the method, each evaluated case was repeated 20 times.}, v-rnk, v-rnk-m\}} \\
    \end{tabular}
    \label{tab:parameters}
\end{table}

\subsection{Implementation}\label{subsec:implementation}

The experimental environment was implemented in Python and based mainly on \textit{NetworkX}~\cite{hagberg2008networkx} package. A part of the developed code, including a spreading model and seed selection methods, was appended to the \textit{Network Diffusion 0.14.0}~\cite{czuba2022networkdiffusion} library. Nonetheless, we stored the main experimental pipeline and scripts to generate and process results in a separate experimental repository.  We also aimed to develop an environment that allows rerunning experiments to reproduce results. Hence, the codebase is published on GitHub (\href{https://github.com/anty-filidor/rank-refined-seeding-bc-infmax-mlnets-ltm/tree/61dd340e40a6889d806c7289b3b2e2df29cfd25a}{www.github.com/anty-filidor/rank-refined-seeding-bc-infmax-mlnets-ltm}) with manuals describing how to run it. Readers who consider enhancing our study with their own seed selection methods or spreading models are welcome to familiarise themselves with the operating principles of the experimental environment's backbone~\cite{czuba2024networkdiffusion}. Regarding the computational resources, the experiments were executed on a workstation with Linux Arch 6.5.3. operating system, Intel Xeon Gold 6238 (x86\_64 architecture) CPU and 394 GB of RAM. The second machine we used had Windows Server 2012R2 installed and was equipped with Intel Xeon E5-2695 (x86\_64 architecture) CPU and 256 GB of RAM.

\section{The First Stage Analysis}\label{sec:results}

As mentioned above, tens of thousands of experiments were conducted during the evaluation. Due to such a large amount of data, finding an appropriate way to process them was necessary. In this chapter, we present step-by-step the methodology applied for processing results, leading to drawing conclusions. Additionally, due to the two-stage nature of the experiments, we first present the results obtained for the base networks. Then, after arranging the evaluated methods in a preliminary ranking, we proceeded to the analysis of experiments conducted on large networks and five seed selection methods, which, in the first phase, were ranked highest. Please note that below we present the aggregated outcomes; however, our GitHub repository contains results of every single experiment we have conducted (including plots like in Fig.~\ref{fig:example_gain} or~\ref{fig:example_diffusion_len} for all of them).

\subsection{Initial Outcomes}\label{subsec:initial_results}

\begin{figure}[ht]
	\centering
	\includegraphics[width=1\linewidth]{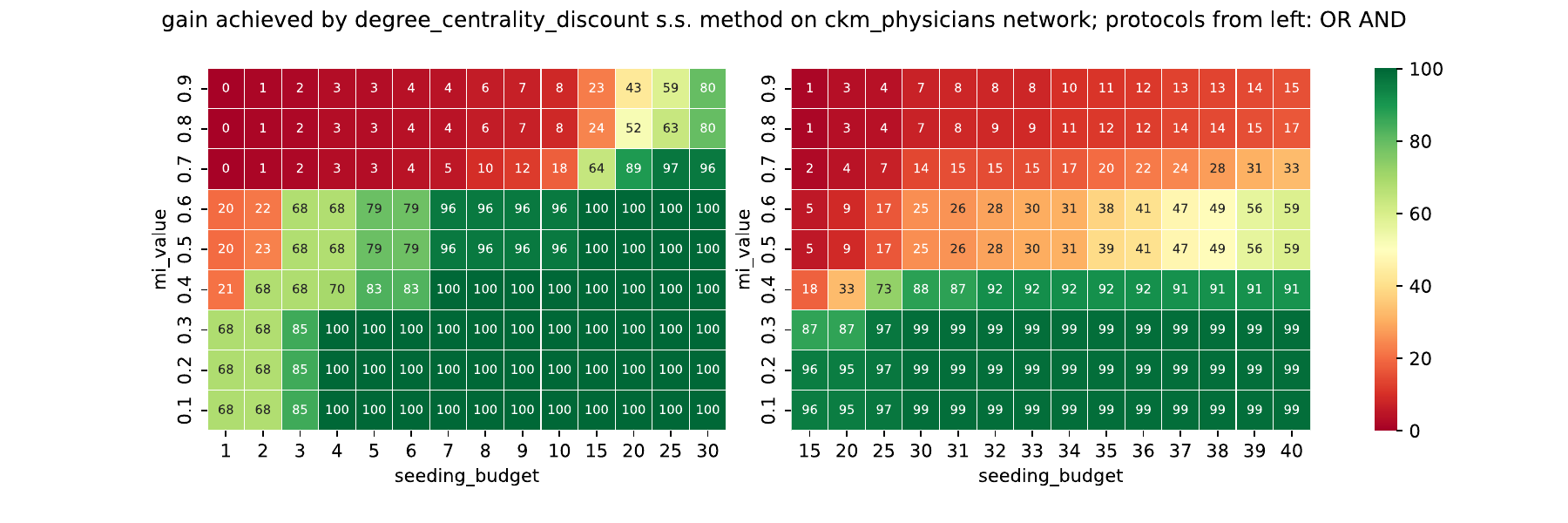}
	\caption{$G$ achieved by \textit{deg-c-d} seed selection method on \textit{ckmp} network. \textbf{Left:} protocol $OR$. \textbf{Right:} protocol $AND$.}
    \label{fig:example_gain}
\end{figure}

\begin{figure}[ht]
	\centering
	\includegraphics[width=1\linewidth]{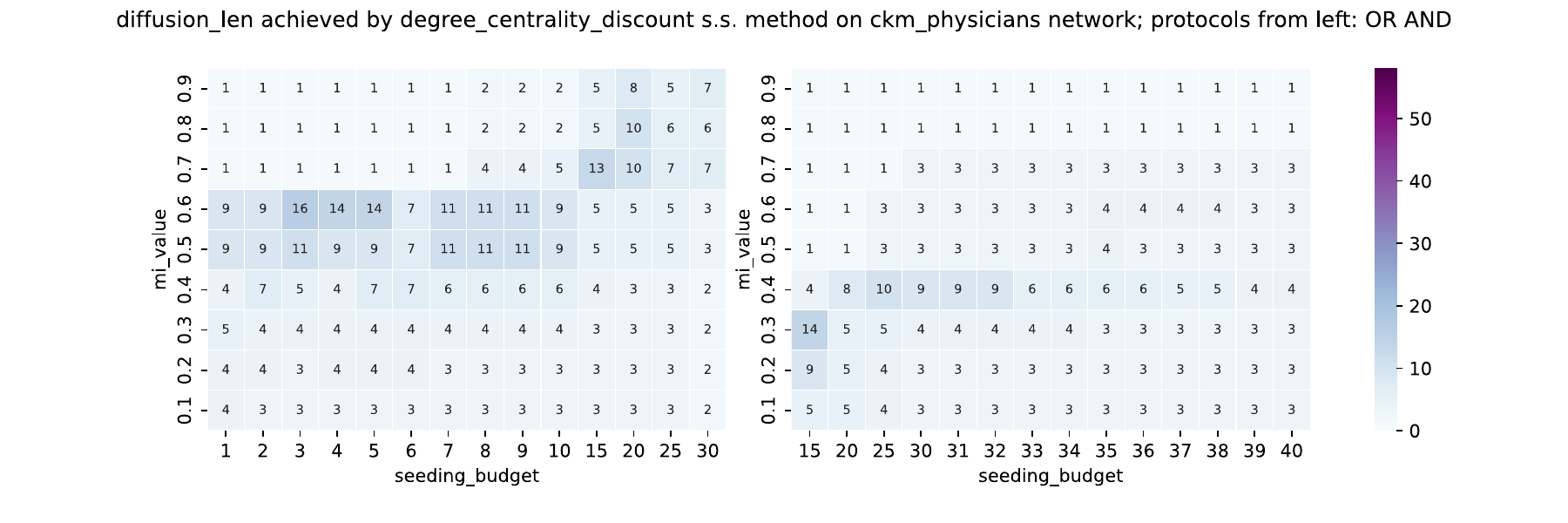}
	\caption{$DL$ achieved by \textit{deg-c-d} seed selection method on \textit{ckmp} network. \textbf{Left:} protocol $OR$. \textbf{Right:} protocol $AND$.}
    \label{fig:example_diffusion_len}
\end{figure}

The experimental pipeline was triggered separately for each seed selection method. From each of the simulations (i.e. an experiment of the MLTM with particular parameters executed on a particular network), we obtained values of $DL$ and $G$. They can be easily visualised on heatmaps, as in Fig.~\ref{fig:example_gain} and Fig.~\ref{fig:example_diffusion_len}.

We analysed them and noticed that their forms differ depending on the applied protocol and seed selection method. However, the most commonly observed pattern was the arrangement of $G$ values into three areas, as in the example image (Fig.~\ref{fig:example_gain}). The first region ("ineffective") covered a subset of parameters where no diffusion occurred or decayed quickly without affecting the network. The next type ("effective") was opposite to the previous one and encompassed a range of parameters where the process was able to activate all nodes in the network. Finally, there was a transitional zone between them where the process spread was already observable but still far from maximal. It is also worth mentioning the relationship between $G$ and $DL$ heatmaps. The maximum diffusion lengths were typically located in the transitional zone and closely matched that area (see Fig.~\ref{fig:example_diffusion_len}). It was also noticeable that $DL$ reached very similar, low values for effective and ineffective combinations of the $\mu$, $s$ parameters.

These dependencies led us to attempt to empirically estimate functions describing observed areas, which is addressed in Sec.~\ref{subsec:curves}. Below, we present a bulk description of the obtained $G$ and $DL$ heatmaps for each type of analysed network.

\subsubsection{Erd\H{o}s-R\'{e}nyi Networks}

We have noticed that the transitional zone is very narrow for random networks. For the protocol $OR$, it does not exist at all, and for the protocol $AND$, it is residual. However, the diffusion length can exceed $50$ epochs, which is a substantial value compared to results obtained for different types of graphs. A larger number of layers does not significantly change the diffusion efficiency, although, not surprisingly, this property helps while protocol $OR$ is used and conversely suppresses diffusion for $AND$. All evaluated seed selection methods perform similarly in such networks, and even random choice produces satisfactory results (which is expected considering how these networks are constructed).

\subsubsection{Scale-free Networks}

In the case of Scale-free networks, the transitional zone is typically present, particularly for the $AND$ strategy, where it occupies a significant area of the map. While that protocol is applied, the process is not able to cover the entire network, regardless of other model parameters and the seed selection method. For protocol $OR$, when the number of layers increases, the transitional area gets reduced, and the overall $G$ grows. Also, for more challenging combinations of $\mu$ and $s$, the MLTM is able to cover the entire graph. Finally, for protocol $OR$, evaluated heuristics are much more efficient than \textit{random}, which is not so noticeable in $AND$ strategy.

\subsubsection{Real Networks}

Real networks revealed the varying effectiveness of the evaluated methods. Furthermore, we observed among them the most significant differences between different implementations of the algorithms (from the perspective of a node and an actor). Therefore, we provide separate descriptions for each of the graphs below.

The transitional region can be clearly observed for the \textit{aucs} graph. For methods that we adapted to the multilayer networks in two ways, both implementations are able to activate the entire graph and yield reasonable results. Generally, the $G$ heatmaps resemble those for the Scale-free networks. Resembling results were obtained for the \textit{ckm-p} graph, although we noticed different behaviour between \textit{k-sh} and \textit{k-sh-m} (in favour of the former). On the other hand, the results for \textit{lazega} network are very similar to Erd\H{o}s-R\'{e}nyi networks --- there is no transitional zone on heatmaps. In this case, we did not observe any anomalies between different approaches for extending seed selection methods either. We noted significantly different results for the \textit{eutr-A} network. Due to its high density, we achieved a coverage close to the maximal for most of the examined model parameters, regardless of the seed selection method applied. However, a spread triggered with \textit{k-sh}, \textit{kpp-sh}, \textit{random} and \textit{v-rnk} did not reveal this phenomenon, where, for the $OR$ protocol, we obtained a high $G$ value regardless of $\mu$ and $s$. Conversely, for the strategy $AND$, the diffusion practically did not occur at all. We also noted that \textit{p-rnk} and \textit{p-rnk-m} were somehow complementary to each other for the protocol $AND$ (i.e. for the sets of parameters in which \textit{p-rnk} yielded high $G$, the \textit{p-rnk-m} was not able to trigger the spread and vice versa).

\subsection{Quantitative Analysis}\label{subsec:quantitative_analysis}

\begin{figure}[ht]
	\centering
	\includegraphics[width=1\linewidth]{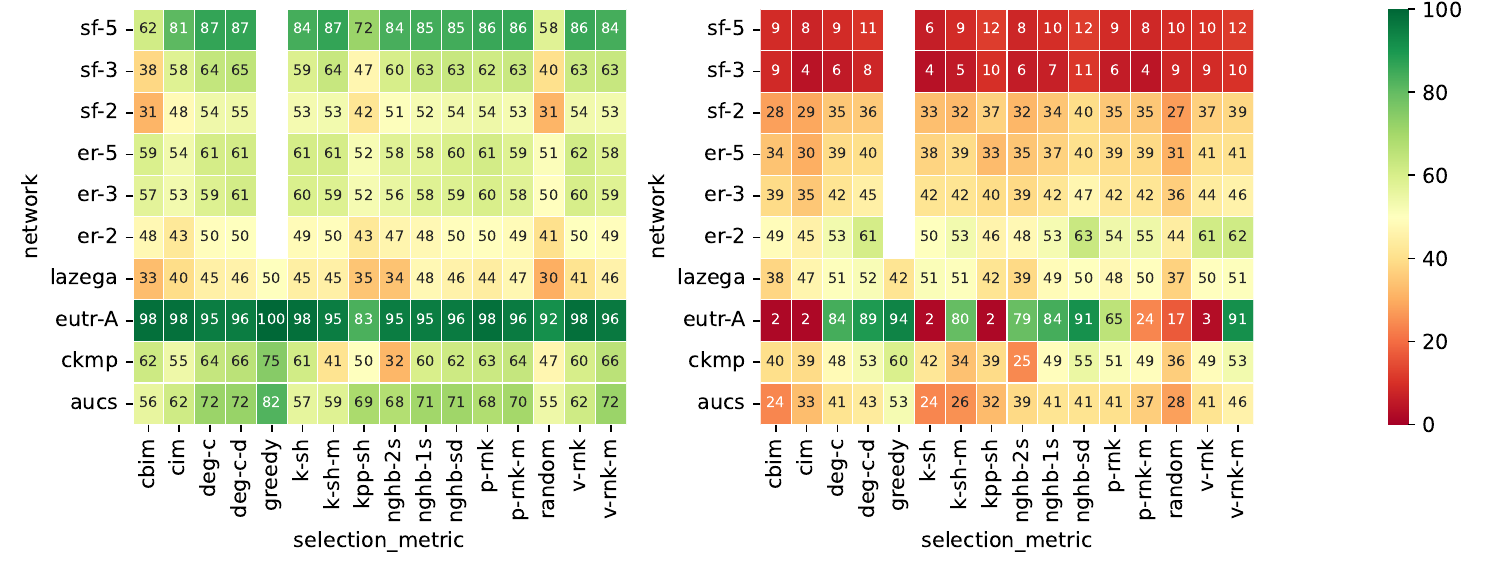}
	\caption{Mean $G$ achieved by seed selection methods on the base 10 networks. \textbf{Left:} protocol $OR$. \textbf{Right:} protocol $AND$.}
    \label{fig:mean_gain}
\end{figure}

\begin{figure}[ht]
	\centering
	\includegraphics[width=1\linewidth]{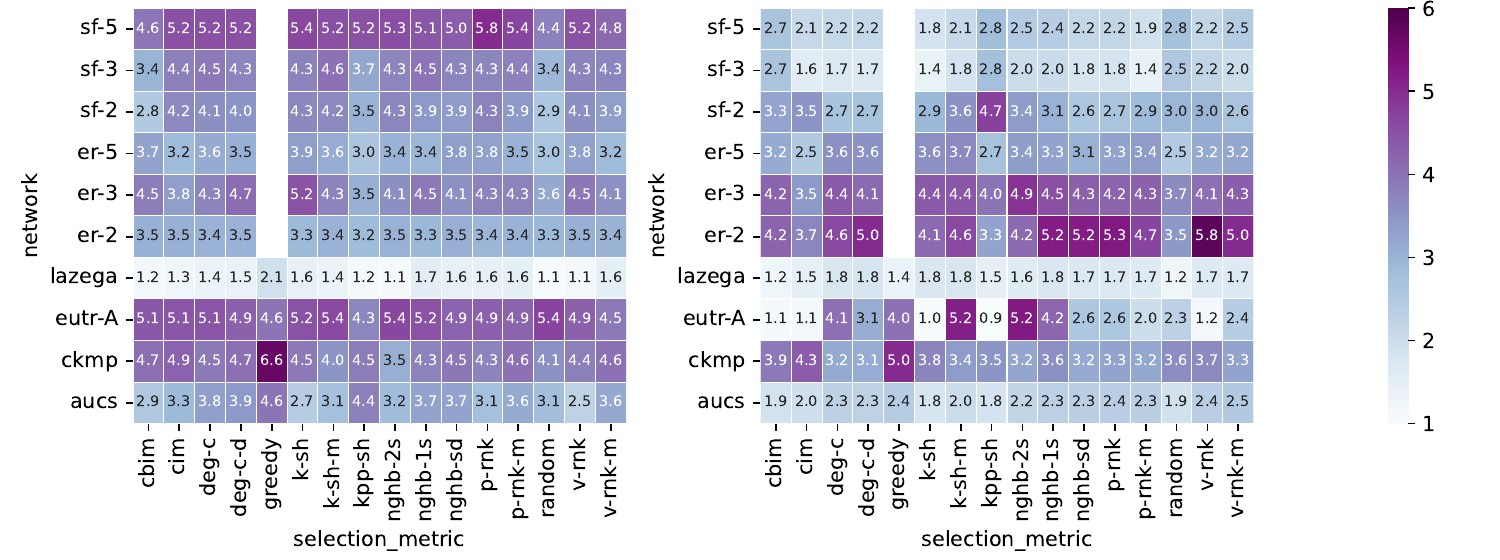}
	\caption{Mean $DL$ achieved by seed selection methods on the base 10 networks. \textbf{Left:} protocol $OR$. \textbf{Right:} protocol $AND$.}
    \label{fig:mean_dl}
\end{figure}

Considering the number of obtained partial results, we aggregated them into a form allowing high-level analysis. Namely, for each combination of: protocol, seed selection method, and network, we computed an average $G$ and $DL$. By that, we could depict results on two-dimensional charts and analyse them independently on the budget size and the activation threshold. Fig.~\ref{fig:mean_dl} and Fig.~\ref{fig:mean_gain} show these heatmaps.

When considering the Fig.~\ref{fig:mean_gain}, it is noteworthy that the average $G$ attained through various seed selection methods exhibits similarity within a specific network. Additionally, correlations between the number of layers and the diffusion efficacy can be observed, as described in Sec.~\ref{subsec:initial_results}. The aggregated results affirm that the network type significantly impacts the achieved $G$, surpassing the influence of seed selection methods.

Regarding average $DL$, it is harder to draw valuable conclusions. That comes from the nature of that metric. The final $G$ is the metric that reveals the efficiency of the spread, and the $DL$ resembles only how long the spreading was present in the network. Nonetheless, Fig.~\ref{fig:mean_dl} shows an average of this value for each network, seed selection method and evaluated protocol of MLTM. Similarly to $G$, we can conclude that the network type strongly affects the $DL$ for the spread triggered with each seed selection method. 



\subsection{Statistical Analysis}\label{subsec:statistics}

\begin{figure}[ht]
	\centering
	\includegraphics[width=1\linewidth]{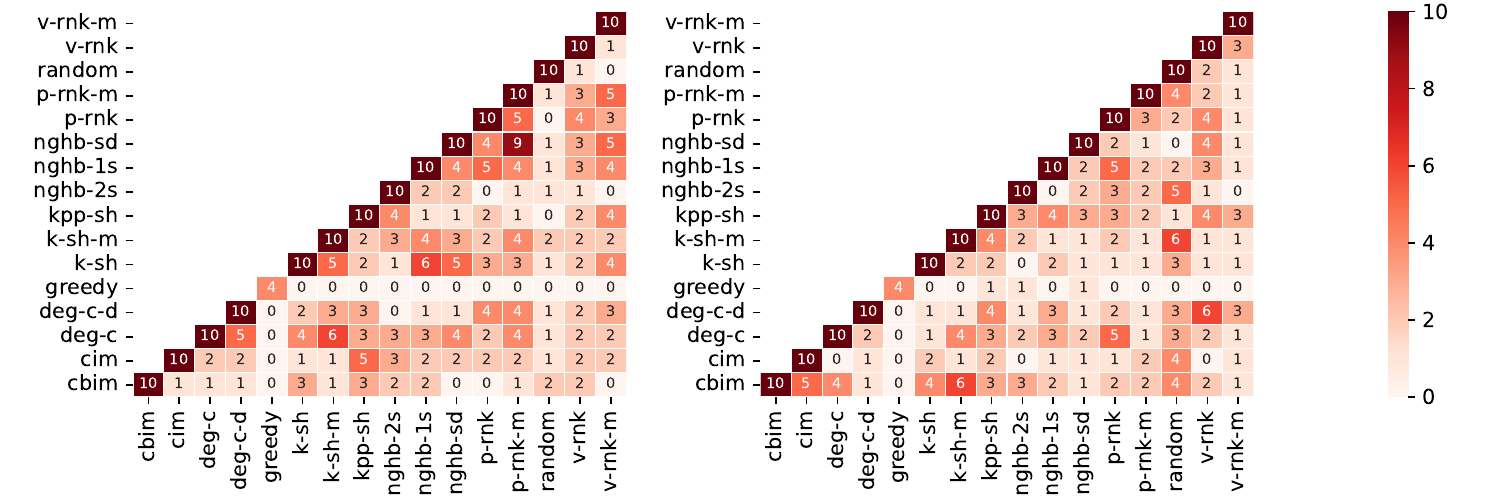}
	\caption{Occurences $p > 0.05$ for Wilcoxon tests performed on series $f(\mu, s) -> G$ aggregated over the base 10 networks. \textbf{Left:} protocol $OR$. \textbf{Right:} protocol $AND$.}
    \label{fig:w_test_all}
\end{figure}

Due to the fact that we obtained comparable outcomes between various seed selection methods regarding the networks used in the experiments, we decided to examine whether the results are statistically similar. Consequently, we employed the Wilcoxon signed rank test for this task. Wilcoxson signed rank test is a non-parametric counterpart of the paired t-test and is often used when one cannot ensure normal distribution of samples~\cite{conover1999practical, hassan2020operational}. Since the obtained results consisted of multiple $0$ and $100$ values of $G$ (i.e. effective and ineffective areas --- see Fig.~\ref{fig:example_gain}), we configured the test to include the zero differences between samples and split them equally between positive and negative ones. The statistics were computed by comparing the aggregated results of $G$ for each seed selection method on each graph to form two data series (one for each protocol). They are depicted on Fig.~\ref{fig:w_test_all}.

Having computed the Wilcoxon test for vectors of $G$ values obtained for seed selection methods on subsequent test cases, one can assess the similarity of how we adapted VoteRank, Page-rank and K-shell Decomposition to multilayer networks. As depicted in Fig.~\ref{fig:w_test_all}, the above methods yielded significantly similar results in less than 50\% of comparisons. That is, results for \textit{k-sh} were similar to \textit{k-sh-m} in five cases for $OR$ and two for $AND$. With regard to \textit{p-rnk} and \textit{p-rnk-m}, they were similar five times for $OR$ and three times for $AND$. Finally, the results for \textit{v-rnk} and \textit{v-rnk-m} attained similarity only one time for $OR$, but three times for $AND$. Hence, it is evident that the adopted variations of these methods are significantly different.

Another interesting conclusion drawn from Fig.~\ref{fig:w_test_all} can be a similarity of the evaluated seed selection methods to \textit{random} and \textit{greedy}. Here, the protocol also plays an important role. Namely, for $OR$ strategy, \textit{random} differs considerably in most comparisons -- it attained similarity to other methods usually one time, probably for the \textit{eutr-A} network, where the spread triggered with all methods attained similarly high $G$ (see Fig.~\ref{fig:mean_gain}). That does not apply to $AND$, where the Wilcoxon test demonstrated some similarities between \textit{random}, especially for \textit{k-sh-m} (six times) and \textit{nghb-2s} (five times). Nonetheless, there are still methods that yield statistically different results than random for most networks: \textit{nghb-sd}, \textit{greedy} (no similarity at all); \textit{kpp-sh}, \textit{v-rnk-m} (one time similar); \textit{nghb-1s}, \textit{p-rnk}, \textit{v-rnk} (two times similar). With regard to the similarity of the seed selection methods to \textit{greedy}, it did not occur at all for the strategy $OR$ and, when the second protocol was applied, only for \textit{kpp-sh}, \textit{ nghb-2s} and \textit{nghb-sd} results were similar in one case.

The last thing we report about the Wilcoxon test is the seed selection methods that yielded statistically similar $G$ for most networks. This is important since methods that yield similar results could be used interchangeably depending on which of them is easier to compute and/or to implement. For protocol $OR$, \textit{p-rnk-m} and \textit{nghb-sd} attained similarity for nine networks, while pairs \textit{k-sh}, \textit{nghb-1s} and \textit{k-sh-m}, \textit{deg-c} for six. Furthermore, for $AND$ strategy, the highest number of similarities was six and occurred between pairs: \textit{cbim}, \textit{k-sh-m} and \textit{deg-c-d}, \textit{v-rnk}, and \textit{k-sh-m}, \textit{random}.

%
%

\subsection{Preliminary Rankings}\label{subsec:all_rankings}

Due to the difficulty in identifying the most effective among the evaluated methods based on average $G$, we decided to transform the results into a more readable form. To achieve this, we calculated rankings based on the $G$ achieved by the spread using each of the seed selection methods. Namely, for each evaluated combination of the MLTM's parameters and seeding budget on a specific network, we computed the rankings of seed selection methods based on the $G$ obtained for spread triggered in such conditions (the greater $G$ indicated a higher place). In such a way, we received a ranking tensor, which shape was $(16, 10, 9, 14, 2)$ (accordingly, number of seed selection methods, number of networks, number of $\mu$ values, number of $s$ values, number of protocols). It was then averaged in the dimensions of $\mu$, $s$ and graph types (real, Scale-free, and Erd\H{o}s-R\'{e}nyi). That resulted in final rankings of seed selection methods. Nonetheless, we chose to retain the distinction due to the different propagation natures between $OR$ and $AND$).

\begin{table}[ht]
\setlength\tabcolsep{4pt}
\begin{subtable}[t]{0.4\textwidth}
    \subcaption{Rankings for protocol $OR$.}
    \begin{tabular}{l||cccc}
        S.S.M. & \multicolumn{4}{c}{Networks} \\
        & Real & Erd\H{o}s-R\'{e}nyi & Scale-free & All \\ \hline \hline
        cbim & 3.29 & 3.51 & \penultimate{5.44} & \thirdtolast{4.00} \\
        cim & 3.15 & \penultimate{4.06} & \thirdtolast{4.81} & 3.92 \\
        deg-c & 2.99 & 2.61 & \third{3.08} & 2.90 \\
        deg-c-d & 2.87 & \third{2.16} & \first{2.80} & \third{2.64} \\
        greedy & \first{1.03} & \multicolumn{2}{c}{not computed} & --- \\
        k-sh & 3.03 & 2.62 & 3.98 & 3.19 \\
        k-sh-m & 3.56 & 2.64 & 3.39 & 3.23 \\
        kpp-sh & \thirdtolast{3.49} & 3.65 & 4.06 & 3.71 \\
        nghb-2s & \penultimate{3.86} & \thirdtolast{3.70} & 4.67 & \penultimate{4.05} \\
        nghb-1s & 3.16 & 2.96 & 3.98 & 3.35 \\
        nghb-sd & 2.97 & 2.58 & 3.67 & 3.06 \\
        p-rnk & \second{2.56} & \second{2.16} & 3.22 & \second{2.64} \\
        p-rnk-m & 2.95 & 2.46 & 3.85 & 3.07 \\
        random & \last{4.15} & \last{4.86} & \last{5.75} & \last{4.84} \\
        v-rnk & 3.01 & \first{1.58} & \second{3.06} & \first{2.59} \\
        v-rnk-m & \third{2.71} & 3.09 & 3.75 & 3.14 \\
    \end{tabular}
    \label{tab:ranking_or_all}
\end{subtable}
\hfill
\begin{subtable}[t]{0.4\textwidth}
    \subcaption{Rankings for protocol $AND$.}
    \begin{tabular}{l||cccc}
        S.S.M. & \multicolumn{4}{c}{Networks} \\
        & Real & Erd\H{o}s-R\'{e}nyi & Scale-free & All \\ \hline \hline
        cbim & \last{7.23} & 5.53 & \thirdtolast{8.31} & \penultimate{7.04} \\
        cim & \thirdtolast{6.54} & \last{6.75} & \last{9.88} & \last{7.60} \\
        deg-c & 3.77 & 4.88 & 6.91 & 5.05 \\
        deg-c-d & 2.62 & 3.28 & \third{4.76} & \third{3.46} \\
        greedy & \first{1.43} & \multicolumn{2}{c}{not computed} & --- \\
        k-sh & 6.52 & 5.15 & \penultimate{9.62} & \thirdtolast{7.04} \\
        k-sh-m & 5.61 & 4.95 & 7.70 & 6.04 \\
        kpp-sh & \penultimate{6.95} & 5.38 & 6.46 & 6.33 \\
        nghb-2s & 5.04 & \thirdtolast{6.17} & 7.76 & 6.19 \\
        nghs-1s & 3.53 & 4.47 & 6.37 & 4.66 \\
        nghb-sd & \third{2.58} & \first{2.67} & \first{2.77} & \first{2.66} \\
        p-rnk & 3.67 & 3.82 & 6.22 & 4.48 \\
        p-rnk-m & 4.59 & 4.22 & 7.98 & 5.50 \\
        random & 6.44 & \penultimate{6.40} & 8.22 & 6.96 \\
        v-rnk & 4.52 & \third{3.21} & 4.98 & 4.27 \\
        v-rnk-m & \second{2.43} & \second{3.05} & \second{3.34} & \second{2.89} \\
    \end{tabular}
    \label{tab:ranking_and_all}
\end{subtable}
\caption{Rankings of the seed selection methods concerning the achieved $G$ for each protocol on the base ten networks, grouped by the graph type. We ignore \textit{greedy} performance in the "All" column because it was computed only for real networks. Greenish cells denote the \first{first}, \second{second}, and \third{third}-ranked seed selection methods, while reddish ones the \thirdtolast{third-to-last}, \penultimate{second-to-last}, and \last{last}.}
\label{tab:ranking_all}
\end{table}

First, in baseline results (Tab.~\ref{tab:ranking_all}) for the $OR$ strategy, we can primarily confirm the initial suspicion that the examined methods shall fall between \textit{random} and \textit{greedy}. For the second evaluated protocol, \textit{greedy} seems to be an upper bound, but there are seed selection methods which are classified below \textit{random}. That is for real networks \textit{cbim}, \textit{kpp-sh}, \textit{cim}, \textit{k-sh}, for Erd\H{o}s-R\'{e}nyi graphs \textit{cim}, for Scale-free networks: \textit{cim}, \textit{k-sh}, and \textit{cbim}. Finally, from the perspective of the bulk average \textit{cim}, \textit{cbim}, and \textit{k-sh}. That finding reveals how much more difficult the influence maximisation problem is in the $AND$ regime compared to the $OR$ strategy.


It is worth noting that, depending on the protocol, the rankings were shifted in different directions. For the $OR$ strategy, the higher places were usually obtained compared to the $AND$. The rankings spanned a smaller range of positions for $OR$ strategy (maximal difference for Erd\H{o}s-R\'{e}nyi graphs between \textit{v-rnk} and \textit{random} --- 3.28) than for $AND$ (maximal difference for Scale-free networks between \textit{nghb-sd} and \textit{cim} --- 7,11). For both cases, these values are not prominent if the total number of the seed selection methods (16 heuristics) is recalled. That indicates that these methods did not significantly stand out from each other.

Having the preliminary rankings, we selected the best methods for evaluation on the large networks. The choice was made based on the "All" columns for both Tab.~\ref{tab:ranking_or_all} and Tab~\ref{tab:ranking_and_all} in a way that we took the top 3 methods, whose union produced the 5 elements set. The final set was obtained from the following methods for protocol $OR$: \textit{v-rnk} (score 2.59), \textit{p-rnk} (score 2.64), and \textit{deg-c-d } (score 2.64). Similarly, for $AND$ from \textit{nghb-sd} (score 2.66), \textit{v-rnk-m} (score 2.89), \textit{deg-c-d} (score 3.46). That is \textit{deg-c-d}, \textit{nghb-sd}, \textit{p-rnk}, \textit{v-rnk}, \textit{v-rnk-m} were chosen for evaluation on the large networks.

\section{The Second Stage Analysis}\label{sec:final_results}

The second stage experiments were conducted on the two largest networks whose properties are described in Tab.~\ref{tab:networks_eda}. The larger one, \textit{timik}, has a higher mean degree (2,87) than \textit{arxiv} (0,17). We chose such different networks intentionally to show the efficiency of the seed selection methods while various conditions occur.

\subsection{Quantitative Analysis}

The initial results revealed heatmaps similar to the example depicted in Fig.~\ref{fig:example_gain} and Fig.~\ref{fig:example_diffusion_len}. However, according to the networks' properties, spread triggered on \textit{timik} yielded much higher $G$ than for \textit{arxiv}. Moreover, for the former graph, the diffusion for the more demanding protocol function --- $AND$ reached maximal gain much more frequently than for the latter graph. We also observed the longest $DL$ while analysing the initial outcomes for that graph --- it reached 95 simulation steps during the experiment with seed selected by \textit{v-rnk} with $s=15$ and MLTM with $\mu=0.6$ and the strategy $AND$. Regarding the \textit{arxiv} network, the final efficiency of the spread depended much more on the protocol applied. In the $OR$ regime, all evaluated seed selection methods yielded similar results (except the deg-c-d, whose effectiveness was lower). Conversely, for protocol $AND$, only seeds determined with \textit{nghb-sd} and \textit{v-rnk-m} caused a high coverage of the graph.

\begin{figure}[ht]
	\centering
	\includegraphics[width=1\linewidth]{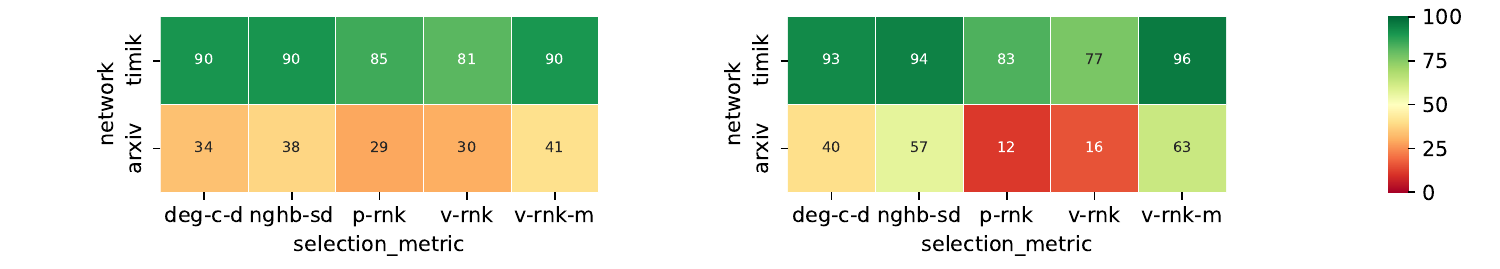}
	\caption{Mean $G$ achieved by seed selection methods on the large networks. \textbf{Left:} protocol $OR$. \textbf{Right:} protocol $AND$.}
    \label{fig:top_mean_gain}
\end{figure}

\begin{figure}[ht]
	\centering
	\includegraphics[width=1\linewidth]{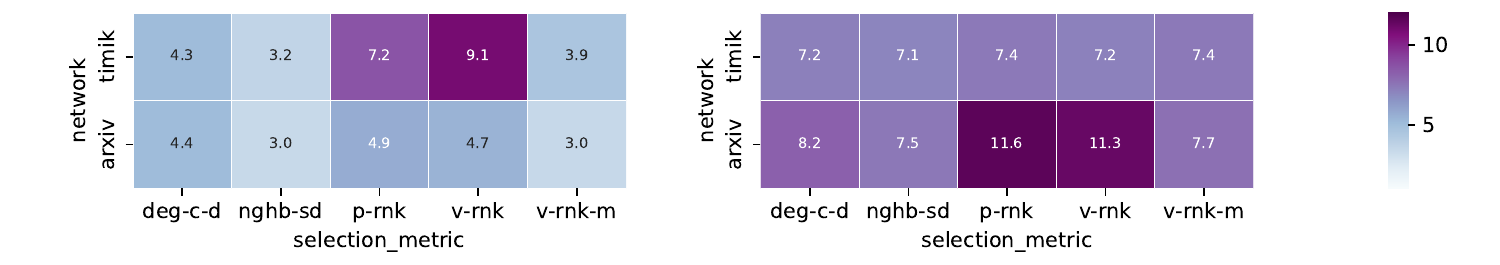}
	\caption{Mean $DL$ achieved by seed selection methods on the large networks. \textbf{Left:} protocol $OR$. \textbf{Right:} protocol $AND$.}
    \label{fig:top_mean_dl}
\end{figure}

That conclusion can be confirmed while heatmaps of the average $G$ and $DL$ are plotted (in a similar way as explained in Sec.~\ref{subsec:quantitative_analysis}). Fig.~\ref{fig:top_mean_gain} and Fig.~\ref{fig:top_mean_dl} reveal that \textit{arxiv} network is much more demanding than \textit{timik}. The amplitude of the average $G$ is bigger for this graph: it reaches 51 ($AND$, \textit{p-rnk}, \textit{v-rnk-m}), while for \textit{timik} only 19 ($AND$, \textit{v-rnk}, \textit{v-rnk-m}). An inverse relation can be observed for the mean $DL$ --- the biggest difference occurred for \textit{timik}: 5.9 ($OR$ \textit{nghb-sd}, \textit{v-rnk}), which was higher than peak for \textit{arxiv}: 4.1 ($AND$ \textit{nghb-sd}, \textit{p-rnk}).

\subsection{Efficiency curves}\label{subsec:curves}

The results obtained during the second stage analyses were also examined for statistical similarities with the Wilcoxon test (as shown in Sec.~\ref{subsec:statistics}). However, none of the pairs constructed from the five seed selection methods demonstrated similarities.

Hence, in order to shed light on the problem from a different perspective, we decided to utilise an observation taken from the raw heatmaps of $G$ (see Fig.~\ref{fig:example_gain}. Namely, we attempted to find functions $\mu = f(s)$ that describe the boundaries between the areas where the diffusion of the MLTM results in heavily different $G$. As denoted in the 
Sec.~\ref{subsec:initial_results}, one can distinguish three regions on such space --- effective (where the process results in the $G$ reaching $100\%$), ineffective (where spreading does not take place), and transitional (where $DL$ is the highest, but achieved $G$ is neither big nor close to $0$). The idea behind this approach is simple --- determining such functions leads to obtaining a general relation between process parameters and $G$. With them (hereinafter called efficiency curves), one can compare the expected $G$ of the diffusion triggered by various seed selection methods with the MLTM's parameters.

\begin{figure}[ht]
     \centering
     \begin{subfigure}[b]{0.23\textwidth}
         \centering
         \includegraphics[width=\textwidth]{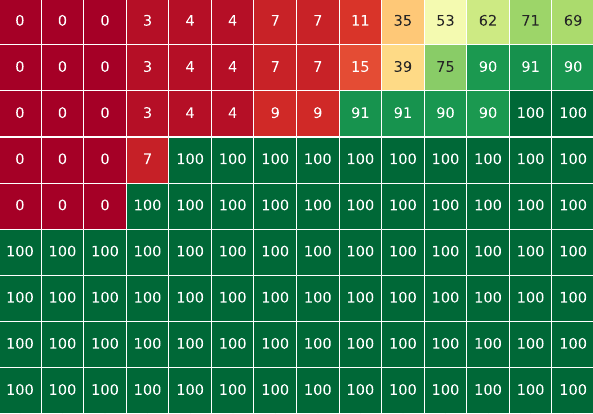}
         \caption{}
         \label{fig:rm_raw_gain}
     \end{subfigure}
     \begin{subfigure}[b]{0.24\textwidth}
         \centering
         \includegraphics[width=\textwidth]{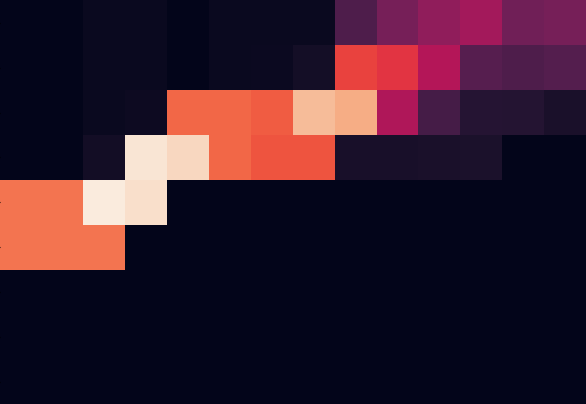}
         \caption{}
         \label{fig:rm_raw_grad}
     \end{subfigure}
     \begin{subfigure}[b]{0.24\textwidth}
         \centering
         \includegraphics[width=\textwidth]{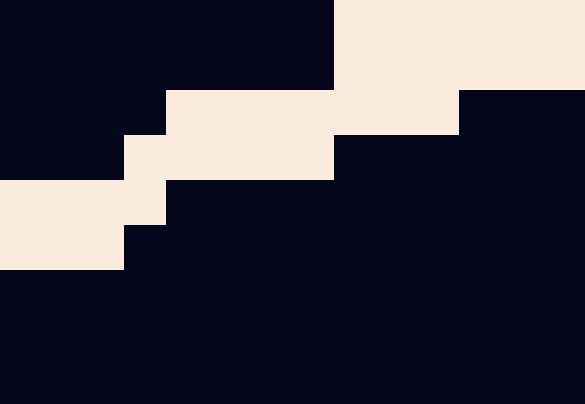}
         \caption{}
         \label{fig:rm_thr_grad}
     \end{subfigure}
          \begin{subfigure}[b]{0.23\textwidth}
         \centering
         \includegraphics[width=\textwidth]{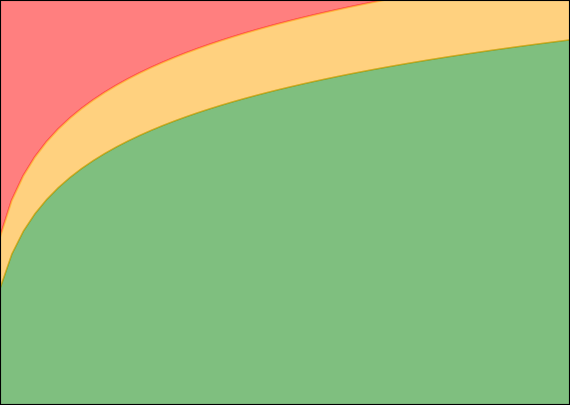}
         \caption{}
         \label{fig:rm_rob_map}
     \end{subfigure}
        \caption{Example of obtaining efficiency curves for the \textit{v-rnk-m} method on \textit{aucs} network with protocol \textit{OR}. Fig.~\ref{fig:rm_raw_gain} shows a heatmap of the raw gain, Fig.~\ref{fig:rm_raw_grad} computed gradient magnitude, Fig.~\ref{fig:rm_thr_grad} thresholded gradient that divides area into three sections, and Fig.~\ref{fig:rm_rob_map} the final curves:: $0.144 \pm 0.013 \cdot log(s) + 0.331 \pm 0.028$ (for efficient-transitional border), $0.155 \pm 0.020 \cdot log(s) + 0.431 \pm 0.044$ (for transitional-inefficient border).}
        \label{fig:rm_example}
\end{figure}

Below, we present a procedure for generating the efficiency curves. At first, we take a raw heatmap depicting relation $f(\mu, s) = G$ (Fig.~\ref{fig:rm_raw_gain}). Then we compute a gradient: $\nabla f = \left[ \frac{\partial G}{\partial s}, \frac{\partial G}{\partial \mu} \right]$ and take its absolute value (Fig.~\ref{fig:rm_raw_grad}). The next step is an application of the threshold — we use $ | \nabla f | > 10$ to filter out small changes of $G$ (Fig.~\ref{fig:rm_thr_grad}). As a result, we expect to obtain a binary matrix with three areas. If they exist, we can determine points of their borders. Finally, we apply a curve fitting procedure on them to obtain an analytical formula for a function sought (Fig.~\ref{fig:rm_rob_map}). We experimented with three of them – a logarithm, a 3rd-degree polynomial and an exponent. The former manifested in the lowest error. Hence, final curves are obtained by fitting $\mu = \alpha \cdot log(s) + \beta$.

Since the efficient-transitional border is more interesting from the perspective of the influence maximisation problem, we decided to depict only approximations of these functions. Nonetheless, in some cases (especially for protocol $AND$), curve fitting errors were too high to treat them as a reliable approximation of the investigated trend. That usually happened when there was a weak border between regions, and we could not determine it properly with gradient methods. However, we decided to show them all to provide a complete report.

\begin{figure}[ht]
	\centering
	\includegraphics[width=1\linewidth]{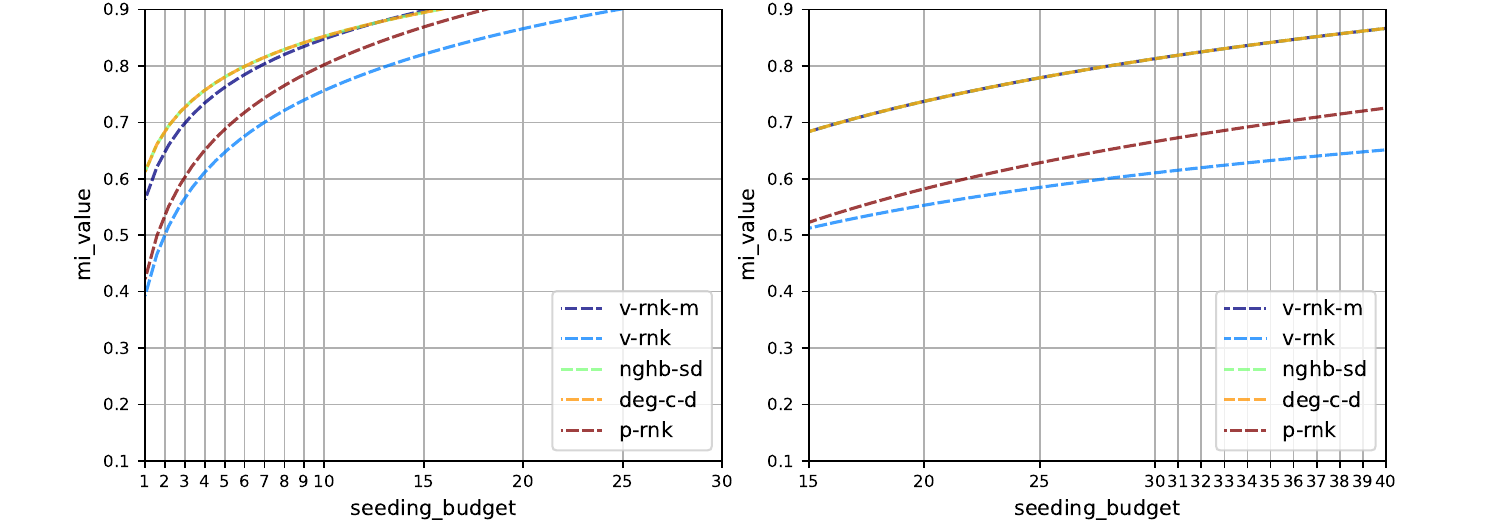}
    \caption{Efficiency curves of the top five seed selection methods computed for \textit{timik} network. \textbf{Left:} protocol $OR$. \textbf{Right:} protocol $AND$. Mean fitting error for $OR$: $\{\alpha=11.42\%, \beta=4.90\%\}$, for $AND$: $\{\alpha=48.10\%, \beta=187.40\%\}$.}
    \label{fig:map_timik}
\end{figure}

\begin{figure}[ht]
	\centering
	\includegraphics[width=1\linewidth]{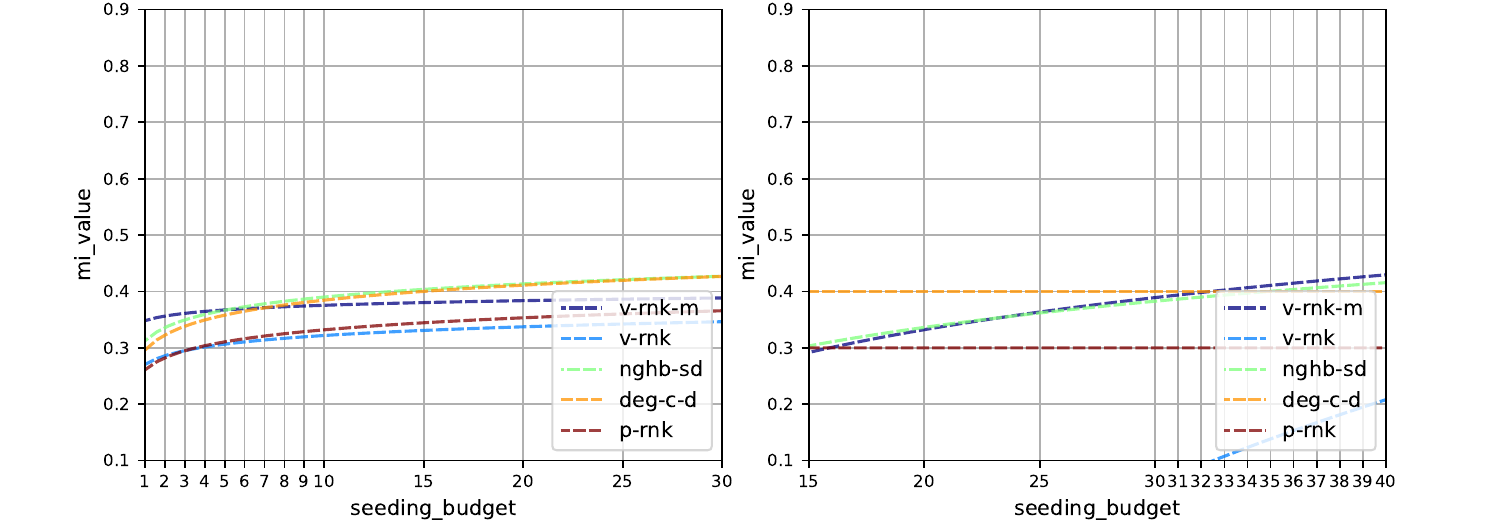}
    \caption{Efficiency curves of the top five seed selection methods computed for \textit{arxiv} network. \textbf{Left:} protocol $OR$. \textbf{Right:} protocol $AND$. Mean fitting error for $OR$: $\{\alpha=46.04\%, \beta=7.00\%\}$, for $AND$: $\{\alpha=5.9E14\%, \beta=7.3E7\%\}.$}
    \label{fig:map_arxiv}
\end{figure}

While analysing Fig.~\ref{fig:map_timik}, one can notice that curves generated for \textit{deg-c-d} and \textit{nghb-s-d} cover each other. That corresponds with Fig.~\ref{fig:top_mean_gain}, where they are very close concerning the mean $G$. The same can be observed between \textit{v-rnk-m}, \textit{nghb-sd}, and \textit{deg-c-d} for the protocol $AND$. 

Curves from Fig.~\ref{fig:map_arxiv} for $OR$ confirm that \textit{arxiv} network is much more demanding --- they are much more flat than their counterparts generated for \textit{timik}. Again, functions obtained for \textit{deg-c-d} and \textit{nghb-s-d} almost cover, which is reflected in Fig.~\ref{fig:top_mean_gain}. The interesting is a case of \textit{v-rnk-m}, which is below the formerly mentioned curves but still attains higher mean $G$ as in Fig.~\ref{fig:top_mean_gain}. Finally, for the $AND$ strategy, approximation errors were too big to draw any valuable conclusions.

It is also valuable to compare the growth of curves in terms of $\mu$ and $s$. The first parameter has a more significant impact on the increase in the performance area. This finding can be successfully utilised to solve practical problems. For instance, in a marketing campaign, it is much better to massively expose the audience to presented content (corresponding to a decrease in $\mu$) than to allocate resources to targeted marketing (corresponding to an increase in $s$). Such an approach allows for much faster coverage of the entire network with the propagated information.

\subsection{Final Rankings}

Similarly to the methodology described in Sec.~\ref{subsec:all_rankings}, we transformed the raw results into the rankings in order to provide the final assessment of the seed selection methods used in the study. They are presented in Tab.~\ref{tab:ranking_top}.

\begin{table}[ht]
    \setlength\tabcolsep{10pt}
    \caption{Rankings of the top five seed selection methods based on the achieved $G$ in the two large networks. Green-shaded cells indicate the \first{first}, \second{second}, and \third{third}-ranked seed selection methods. The rightmost column presents the computational complexity of the given method.}
    \begin{tabular}{l||ccc|c}
    S.S.M. & $OR$ & $AND$ & Avg. & Computational complexity \\ \hline \hline
    deg-c-d & 3.44 & \third{3.27} & \third{3.36} & $O(|A| + |E| + |A|^2|L|)$ \\
    nghb-sd & \second{2.76} & \second{2.14} & \second{2.45} & $O(|A| + |E| + |A|^2|L|)$ \\
    p-rnk & \third{2.77} & 4.13 & 3.45 & $O(|N| + |L|(1 + |A| + |E| + |N|) + |A|log|A|)$ \\
    v-rnk & 2.77 & 4.24 & 3.51 & $O(|N| + |L|(1 + |A| + |E| + |N|log|N| + |E|^2 / |N|) + |A|log|A|)$\\
    v-rnk-m & \first{2.13} & \first{1.21} & \first{1.67} &  $O(|A||E| + |A|^2 + |E|^2 / |A|)$ \\
    \end{tabular}
    \label{tab:ranking_top}
\end{table}

The results reveal a dominance of \textit{v-rnk-m}, the highest classified method for both protocols and on average. The second one was \textit{nghb-sd}, which also took the second place for all criteria. Finally, the podium was closed by \textit{deg-c-d}, which was ousted by \textit{p-rnk} for $OR$ strategy. Tab.~\ref{tab:ranking_top} shows another interesting fact. Namely, that the relatively simple heuristics --- \textit{nghb-sd} and \textit{deg-c-d} outperformed much more sophisticated \textit{p-rank} or \textit{v-rnk}.

However, let us refer to the efficiency curves (Sec.~\ref{subsec:curves}) while analysing the final ranking. In that case, one can speculate that the critical factor determining the general efficiency of the seed selection method is its behaviour in the demanding environment. In other words, a performance under conditions like parameters determining the transitional area, the $AND$ strategy, or if the sparse network is used as a medium for the propagation. Indeed, such circumstances reveal the differences between seed selection methods and make it easier to assess them.

Although this work is not focused on efficient implementations of the employed seed selection methods, we shortly discuss the complexity of those that demonstrate the highest effectiveness, even in their prototype implementations. Tab.~\ref{tab:ranking_top} reveals that the most effective methods generally involve more computationally demanding procedures. Before comparing complexities, let us note that in typical multilayer networks, the number of layers ($|L|$) is usually small, and connections are sparse, meaning $|N|$ and $|E|$ are of the same order of magnitude. The top-ranked \textit{v-rnk-m} requires intensive updates to voting scores and iterative selections, leading to a considerably higher computational burden. The two discounting methods, \textit{nghb-sd} and \textit{deg-c-d}, which ranked second and third, also exhibit a quadratic scaling trend. In contrast, methods based on averaging centralities computed for each actor (\textit{p-rnk} and \textit{v-rnk}) scale more favourably yet demonstrate the lowest efficiency in terms of performance. These differences highlight an inherent trade-off: while more sophisticated approaches may achieve higher effectiveness, their computational cost remains a critical factor, particularly when scaling to large networks.

\section{Conclusions}\label{sec:conclusion}

In this study, we presented an evaluation of various centrality-based seed selection methods for the Budget Constrained Influence Maximisation problem in multilayer networks under the Multilayer Linear Threshold Model.

The first aspect of our work involves proposing the generalisation of three centrality functions (K-shell Decomposition~\cite{shai2007kshell}, PageRank~\cite{page1999pagerank}, and VoteRank~\cite{zhang2016identifying}) for multilayer networks. They can be further explored in studies focused on centrality in multilayer networks, as these metrics were originally designed to evaluate the centrality of agents (their application to the problem of influence maximisation represents only an alternative way to leverage their capabilities). The next group of methods like Community Based Influence Maximisation~\cite{Venkatakrishna2022CBIM}, Clique Influence Maximisation~\cite{Venkatakrishna2022CIM} and K++ Shell~\cite{Venkatakrishna2023KppShell}, although originally defined for multilayer networks, were adapted by us to the problem of the influence maximisation based on actors, not nodes. We also introduced a new heuristic --- Neighbourhood Size Discount. Finally, all of them (plus a greedy routine and a random choice as hypothetical boundaries) were evaluated.

Our additional contribution is an examination of influence under the MLTM by proving that it is not submodular. We also proposed a new approach to measure it --- $G$, which (in contrast to the classic $\sigma$) considers the seed set size. 

The next area of our work focuses on assessing the effectiveness of the aforementioned heuristics, both quantitatively and qualitatively. The comprehensive evaluation offered insights into their performance, contributing to understanding their applicability. Apart from $G$, we developed another way to assess visually the performance of the seed selection methods in the form of the effectiveness curves. They determine a boundary between the parameters where diffusion effectiveness decreases and where the spread is maximal.

We can conclude that all the evaluated parameters, including the internal parameters of the MLTM, the network type, the budget, and the seed selection method, collectively contribute to the overall effectiveness of influence maximisation. Moreover, there is no seed selection method which always provides the best results. Except for the random choice and greedy approach, which stand out from other results, the mean performance of evaluated methods (Fig.\ref{fig:mean_gain}) is levelled for protocol $OR$. Interestingly, the same phenomena cannot be observed for more rigorous $AND$, where the mean $G$ is more scattered among seed selection methods. Therefore, in real-world situations (such as opinion propagation), it may sometimes be more effective to focus on reducing the activation threshold (e.g., through mass propaganda or promoting desired attitudes in mass media) rather than searching for a specific set of "influencers". Moreover, when given methods are ranked relatively close to each other, one can consider a criterium of time complexity to choose the proper algorithm so that fast yet akin in results routine is preferred (e.g. lightweight Degree Centrality compared to a heavier PageRank).

The preliminary assessment performed on ten networks revealed that the five methods yielding the best results are Degree Centrality Discount, Neighbourhood Size Discount, PageRank (variant \textit{p-rnk}), and VoteRank (both variants: \textit{v-rnk}, \textit{v-rnk-m}). Finally, the second stage analysis made with the large graphs and heuristics mentioned above surfaced \textit{v-rnk-m} as the best one. A production-ready implementation of all these methods can be found in the Network Diffusion~\cite{czuba2024networkdiffusion} library.

Regarding this study's limitations, it is essential to mention that we analysed only seed selection methods from the rank refined subcategory of the heuristics group (according to~\cite{singh2022influence}). Moreover, we considered computational complexity only for the most effective methods. However, since all the examined approaches were developed specifically for this work, optimising their implementation for efficiency was not our primary objective. Consequently, the scalability of the employed seed selection methods remains an open avenue for future research and practical applications. Additionally, this work does not explore machine learning approaches recently proposed for single-layer networks. Additionally, we limited the threshold in the spreading model to be homogeneous among actors to limit additional degrees of freedom in the study. Both of these issues can be addressed in future research.

As mentioned, ranking-based seed selection methods do not require knowledge of the propagation dynamics, making their applicability broader. Hence, it could be interesting to consider them in problems related to influence maximisation and compare to the state-of-the-art methods, like for diffusion source inferring~\cite{liu2025tnnls_2023_3321767, liu2024tnet_2024_3382546} or diffusion containment~\cite{liu2024tifs_2023_3338423, liu2024tnse_2024_3388994} which are still fields open to the exploration for multilayer networks. The next interesting topic can be addressing the main limitation of the heuristics we used --- a requirement of access to the entire network. That could be done by studies oriented on generalising the machine learning methods for seed selection.

\backmatter

\bmhead{Competing interest} None.

\bmhead{Supplementary information}

The codebase is published on GitHub (\href{https://github.com/anty-filidor/rank-refined-seeding-bc-infmax-mlnets-ltm/tree/61dd340e40a6889d806c7289b3b2e2df29cfd25a}{www.github.com/anty-filidor/rank-refined-seeding-bc-infmax-mlnets-ltm}) together with manuals describing how to run it, as well as all detailed results for each experiment and simulation we have executed.

\bmhead{Acknowledgements}

This research was partially supported by the National Science Centre, Poland, under Grant no. 2022/45/B/ST6/04145, the Polish Ministry of Education and Science within the programme “International Projects Co-Funded”, and the EU under the Horizon Europe, grant no. 101086321 OMINO. Views and opinions expressed are, however, those of the authors only and do not necessarily reflect those of the National Science Centre, Polish Ministry of Education and Science, EU or the European Research Executive Agency.

\bibliography{references.bib}

\begin{thebibliography}{10}
\providecommand{\url}[1]{{#1}}
\providecommand{\urlprefix}{URL }
\providecommand{\doi}[1]{\url{https://doi.org/#1}}
\bibcommenthead

\bibitem{kempe2003maximizing}
D.~Kempe, J.~Kleinberg, {\'E}.~Tardos, \emph{Maximizing the Spread of Influence Through a Social Network}, in \emph{9th ACM SIGKDD international conference on Knowledge discovery and data mining} (2003), pp. 137--146.
\newblock \doi{10.1145/956750.956769}.
\newblock \urlprefix\url{https://doi.org/10.1145/956750.956769}

\bibitem{singh2022influence}
S.S. Singh, D.~Srivastva, M.~Verma, J.~Singh, Influence maximization frameworks, performance, challenges and directions on social network: A theoretical study.
\newblock Journal of King Saud University-Computer and Information Sciences \textbf{34}(9), 7570--7603 (2022).
\newblock \doi{https://doi.org/10.1016/j.jksuci.2021.08.009}.
\newblock \urlprefix\url{https://www.sciencedirect.com/science/article/pii/S1319157821002123}

\bibitem{leskovec2007cost}
J.~Leskovec, A.~Krause, C.~Guestrin, C.~Faloutsos, J.~VanBriesen, N.~Glance, \emph{Cost-effective outbreak detection in networks}, in \emph{Proceedings of the 13th ACM SIGKDD International Conference on Knowledge Discovery and Data Mining} (Association for Computing Machinery, New York, NY, USA, 2007), KDD '07, p. 420–429.
\newblock \doi{10.1145/1281192.1281239}.
\newblock \urlprefix\url{https://doi.org/10.1145/1281192.1281239}

\bibitem{wang2010communitybasedgreedy}
Y.~Wang, G.~Cong, G.~Song, K.~Xie, \emph{Community-based greedy algorithm for mining top-K influential nodes in mobile social networks}, in \emph{Proceedings of the 16th ACM SIGKDD International Conference on Knowledge Discovery and Data Mining} (Association for Computing Machinery, New York, NY, USA, 2010), KDD '10, p. 1039–1048.
\newblock \doi{10.1145/1835804.1835935}.
\newblock \urlprefix\url{https://doi.org/10.1145/1835804.1835935}

\bibitem{sheikhahmadi2015improving}
A.~Sheikhahmadi, M.A. Nematbakhsh, A.~Shokrollahi, Improving detection of influential nodes in complex networks.
\newblock Physica A: Statistical Mechanics and its Applications \textbf{436}, 833--845 (2015).
\newblock \doi{10.1016/j.physa.2015.04.035}.
\newblock \urlprefix\url{https://www.sciencedirect.com/science/article/pii/S0378437115004033}

\bibitem{wang2016effective}
X.~Wang, Y.~Su, C.~Zhao, D.~Yi, Effective identification of multiple influential spreaders by degreepunishment.
\newblock Physica A: Statistical Mechanics and its Applications \textbf{461}, 238--247 (2016).
\newblock \doi{https://doi.org/10.1016/j.physa.2016.05.020}.
\newblock \urlprefix\url{https://www.sciencedirect.com/science/article/pii/S0378437116302059}

\bibitem{zhang2016identifying}
J.X. Zhang, D.B. Chen, Q.~Dong, Z.D. Zhao, Identifying a set of influential spreaders in complex networks.
\newblock Scientific reports \textbf{6}(1), 27823 (2016).
\newblock \doi{10.1038/srep27823}.
\newblock \urlprefix\url{https://doi.org/10.1038/srep27823}

\bibitem{he2015novel}
J.L. He, Y.~Fu, D.B. Chen, A novel top-k strategy for influence maximization in complex networks with community structure.
\newblock PloS one \textbf{10}(12), e0145283 (2015).
\newblock \doi{10.1371/journal.pone.0162066}.
\newblock \urlprefix\url{https://doi.org/10.1371/journal.pone.0162066}

\bibitem{bao2017identifying}
Z.K. Bao, J.G. Liu, H.F. Zhang, Identifying multiple influential spreaders by a heuristic clustering algorithm.
\newblock Physics Letters A \textbf{381}(11), 976--983 (2017).
\newblock \doi{10.1016/j.physleta.2017.01.043}.
\newblock \urlprefix\url{https://www.sciencedirect.com/science/article/pii/S0375960116309379}

\bibitem{sadaf2022maximising}
A.~Sadaf, L.~Mathieson, P.~Br{\'o}dka, K.~Musial, Maximising influence spread in complex networks by utilising community-based driver nodes as seeds.
\newblock Information Management and Big Data pp. 126--141 (2023).
\newblock \doi{10.1007/978-3-031-35445-8_10}.
\newblock \urlprefix\url{https://doi.org/10.1007/978-3-031-35445-8_10}

\bibitem{kimura2006tracable}
M.~Kimura, K.~Saito, \emph{Tractable Models for Information Diffusion in Social Networks}, in \emph{Knowledge Discovery in Databases: PKDD 2006}, ed. by J.~F{\"u}rnkranz, T.~Scheffer, M.~Spiliopoulou (Springer Berlin Heidelberg, Berlin, Heidelberg, 2006), pp. 259--271.
\newblock \doi{10.1007/11871637_27}.
\newblock \urlprefix\url{https://link.springer.com/chapter/10.1007/11871637_27}

\bibitem{cheng2010ScalableIM}
W.~Chen, Y.~Yuan, L.~Zhang, \emph{Scalable Influence Maximization in Social Networks under the Linear Threshold Model}, in \emph{2010 IEEE International Conference on Data Mining} (2010), pp. 88--97.
\newblock \doi{10.1109/ICDM.2010.118}.
\newblock \urlprefix\url{https://ieeexplore.ieee.org/document/5693962}

\bibitem{kim2013scalable}
J.~Kim, S.K. Kim, H.~Yu, \emph{Scalable and parallelizable processing of influence maximization for large-scale social networks?}, in \emph{2013 IEEE 29th International Conference on Data Engineering (ICDE)} (2013), pp. 266--277.
\newblock \doi{10.1109/ICDE.2013.6544831}.
\newblock \urlprefix\url{https://ieeexplore.ieee.org/document/6544831}

\bibitem{cheng2013staticgreedy}
S.~Cheng, H.~Shen, J.~Huang, G.~Zhang, X.~Cheng, \emph{StaticGreedy: solving the scalability-accuracy dilemma in influence maximization}, in \emph{Proceedings of the 22nd ACM International Conference on Information \& Knowledge Management} (Association for Computing Machinery, New York, NY, USA, 2013), CIKM '13, pp. 509--518.
\newblock \doi{10.1145/2505515.2505541}.
\newblock \urlprefix\url{https://doi.org/10.1145/2505515.2505541}

\bibitem{zhou2015upper}
C.~Zhou, P.~Zhang, W.~Zang, L.~Guo, On the upper bounds of spread for greedy algorithms in social network influence maximization.
\newblock IEEE Transactions on Knowledge and Data Engineering \textbf{27}(10), 2770--2783 (2015).
\newblock \doi{10.1109/TKDE.2015.2419659}.
\newblock \urlprefix\url{https://ieeexplore.ieee.org/document/7079460}

\bibitem{borgs2014reversereachablesets}
C.~Borgs, M.~Brautbar, J.~Chayes, B.~Lucier, \emph{Maximizing social influence in nearly optimal time}, in \emph{Proceedings of the Twenty-Fifth Annual ACM-SIAM Symposium on Discrete Algorithms} (Society for Industrial and Applied Mathematics, USA, 2014), SODA '14, p. 946–957.
\newblock \doi{10.5555/2634074.2634144}.
\newblock \urlprefix\url{https://dl.acm.org/doi/10.5555/2634074.2634144}

\bibitem{Sun2021InfMaxRR}
G.~Sun, C.C. Chen, {Influence Maximization Algorithm Based on Reverse Reachable Set}.
\newblock Mathematical Problems in Engineering \textbf{2021} (2021).
\newblock \doi{10.1155/2021/5535843}.
\newblock \urlprefix\url{https://www.hindawi.com/journals/mpe/2021/5535843/}

\bibitem{tiukhova2022influencer}
E.~Tiukhova, E.~Penaloza, M.~Óskarsdóttir, H.~Garcia, A.C. Bahnsen, B.~Baesens, M.~Snoeck, C.~Bravo.
\newblock Influencer detection with dynamic graph neural networks (2022).
\newblock \doi{10.48550/arXiv.2307.08131}.
\newblock \urlprefix\url{https://arxiv.org/abs/2307.08131}

\bibitem{kou2023identify}
J.~Kou, P.~Jia, J.~Liu, J.~Dai, H.~Luo, Identify influential nodes in social networks with graph multi-head attention regression model.
\newblock Neurocomputing \textbf{530}, 23--36 (2023).
\newblock \doi{https://doi.org/10.1016/j.neucom.2023.01.078}.
\newblock \urlprefix\url{https://www.sciencedirect.com/science/article/pii/S0925231223001108}

\bibitem{hajarathaiah2022generalization}
K.~Hajarathaiah, M.K. Enduri, S.~Dhuli, S.~Anamalamudi, L.R. Cenkeramaddi, Generalization of relative change in a centrality measure to identify vital nodes in complex networks.
\newblock IEEE Access \textbf{11}, 808--824 (2023).
\newblock \doi{10.1109/ACCESS.2022.3232288}.
\newblock \urlprefix\url{https://ieeexplore.ieee.org/document/9999215}

\bibitem{rezaei2023machine}
A.~{Asgharian Rezaei}, J.~Munoz, M.~Jalili, H.~Khayyam, A machine learning-based approach for vital node identification in complex networks.
\newblock Expert Systems with Applications \textbf{214}, 119086 (2023).
\newblock \doi{https://doi.org/10.1016/j.eswa.2022.119086}.
\newblock \urlprefix\url{https://www.sciencedirect.com/science/article/pii/S0957417422021042}

\bibitem{bucur2020top}
D.~Bucur, Top influencers can be identified universally by combining classical centralities.
\newblock Scientific reports \textbf{10}(1), 1--14 (2020).
\newblock \doi{10.1038/s41598-020-77536-7}.
\newblock \urlprefix\url{https://www.nature.com/articles/s41598-020-77536-7}

\bibitem{chen2023deepim}
C.~Ling, J.~Jiang, J.~Wang, M.T. Thai, L.~Xue, J.~Song, M.~Qiu, L.~Zhao, \emph{Deep graph representation learning and optimization for influence maximization}, in \emph{Proceedings of the 40th International Conference on Machine Learning} (JMLR.org, 2023), ICML'23.
\newblock \doi{10.5555/3618408.3619288}.
\newblock \urlprefix\url{https://proceedings.mlr.press/v202/ling23b/ling23b.pdf}

\bibitem{brodka2021sequential}
P.~Br\'{o}dka, J.~Jankowski, R.~Michalski, Sequential seeding in multilayer networks.
\newblock Chaos: An Interdisciplinary Journal of Nonlinear Science \textbf{31}(3), 033130 (2021).
\newblock \doi{10.1063/5.0023427}.
\newblock \urlprefix\url{https://doi.org/10.1063/5.0023427}.
\newblock {\href{https://arxiv.org/abs/https://doi.org/10.1063/5.0023427}{{https://doi.org/10.1063/5.0023427}}}

\bibitem{jankowski2017balancing}
J.~Jankowski, P.~Br{\ifmmode\acute{o}\else\'{o}\fi}dka, P.~Kazienko, B.K. Szymanski, R.~Michalski, T.~Kajdanowicz, {Balancing Speed and Coverage by Sequential Seeding in Complex Networks}.
\newblock Scientific Reports \textbf{7}(891), 1--11 (2017).
\newblock \doi{10.1038/s41598-017-00937-8}.
\newblock \urlprefix\url{https://www.nature.com/articles/s41598-017-00937-8}

\bibitem{seeman2013adaptive}
L.~Seeman, Y.~Singer, \emph{Adaptive Seeding in Social Networks}, in \emph{2013 IEEE 54th Annual Symposium on Foundations of Computer Science} (2013), pp. 459--468.
\newblock \doi{10.1109/FOCS.2013.56}.
\newblock \urlprefix\url{https://ieeexplore.ieee.org/document/6686182}

\bibitem{goldenberg2018timing}
D.~Goldenberg, A.~Sela, E.~Shmueli, Timing matters: Influence maximization in social networks through scheduled seeding.
\newblock IEEE Transactions on Computational Social Systems \textbf{5}(3), 621--638 (2018).
\newblock \doi{10.1109/TCSS.2018.2852742}.
\newblock \urlprefix\url{https://ieeexplore.ieee.org/document/8424548}

\bibitem{sela2018active}
A.~Sela, D.~Goldenberg, I.~Ben-Gal, E.~Shmueli, Active viral marketing: Incorporating continuous active seeding efforts into the diffusion model.
\newblock Expert Systems with Applications \textbf{107}, 45--60 (2018).
\newblock \doi{https://doi.org/10.1016/j.eswa.2018.04.016}.
\newblock \urlprefix\url{https://www.sciencedirect.com/science/article/pii/S095741741830246X}

\bibitem{kitsak2010superspreaders}
M.~Kitsak, L.K. Gallos, S.~Havlin, F.~Liljeros, L.~Muchnik, H.E. Stanley, H.A. Makse, {Identification of influential spreaders in complex networks}.
\newblock Nature Phys \textbf{6}, 888--893 (2010).
\newblock \doi{10.1038/nphys1746}.
\newblock \urlprefix\url{https://doi.org/10.1038/nphys1746}

\bibitem{SVIRIDENKO2004Knapsack}
M.~Sviridenko, A note on maximizing a submodular set function subject to a knapsack constraint.
\newblock Operations Research Letters \textbf{32}(1), 41--43 (2004).
\newblock \doi{https://doi.org/10.1016/S0167-6377(03)00062-2}.
\newblock \urlprefix\url{https://www.sciencedirect.com/science/article/pii/S0167637703000622}

\bibitem{yuan2024gbim}
Z.~Yuan, M.~Shao, Z.~Chen, {Graph Bayesian Optimization for Multiplex Influence Maximization}.
\newblock AAAI \textbf{38}(20), 22475--22483 (2024).
\newblock \doi{10.1609/aaai.v38i20.30255}.
\newblock \urlprefix\url{https://doi.org/10.1609/aaai.v38i20.30255}

\bibitem{michalski2014temporal}
R.~Michalski, T.~Kajdanowicz, P.~Br{\'o}dka, P.~Kazienko, Seed selection for spread of influence in social networks: Temporal vs. static approach.
\newblock New Generation Computing \textbf{32}(3), 213--235 (2014).
\newblock \doi{10.1007/s00354-014-0402-9}.
\newblock \urlprefix\url{https://doi.org/10.1007/s00354-014-0402-9}

\bibitem{liu2024rsif_2023_0625}
Y.~Liu, X.~Wang, X.~Wang, L.~Yan, S.~Zhao, Z.~Wang, {Individual-centralized seeding strategy for influence maximization in information-limited networks}.
\newblock J. R. Soc. Interface \textbf{21}(214) (2024).
\newblock \doi{10.1098/rsif.2023.0625}.
\newblock \urlprefix\url{https://royalsocietypublishing.org/doi/10.1098/rsif.2023.0625}

\bibitem{hpv2024usgov}
U.D. of~Health \& Human~Services.
\newblock {HPV Infection} (2024).
\newblock \urlprefix\url{https://www.cdc.gov/hpv/parents/about-hpv.html}.
\newblock [Online; accessed 18. May 2024]

\bibitem{wei2012profitmaximisationltm}
W.~Lu, L.V. Lakshmanan, \emph{Profit Maximization over Social Networks}, in \emph{2012 IEEE 12th International Conference on Data Mining} (2012), pp. 479--488.
\newblock \doi{10.1109/ICDM.2012.145}.
\newblock \urlprefix\url{https://ieeexplore.ieee.org/document/6413877}

\bibitem{tang2018profitmaximsation}
J.~Tang, X.~Tang, J.~Yuan, \emph{Towards Profit Maximization for Online Social Network Providers}, in \emph{IEEE INFOCOM 2018 - IEEE Conference on Computer Communications} (2018), pp. 1178--1186.
\newblock \doi{10.1109/INFOCOM.2018.8485975}.
\newblock \urlprefix\url{https://ieeexplore.ieee.org/document/8485975}

\bibitem{tang2015influence}
Y.~Tang, Y.~Shi, X.~Xiao, \emph{Influence maximization in near-linear time: A martingale approach}, in \emph{Proceedings of the 2015 ACM SIGMOD international conference on management of data} (2015), pp. 1539--1554.
\newblock \doi{10.1145/2723372.2723734}.
\newblock \urlprefix\url{https://doi.org/10.1145/2723372.2723734}

\bibitem{watroba2023influence}
P.~W{\k{a}}troba, P.~Br{\'o}dka, Influence of information blocking on the spread of virus in multilayer networks.
\newblock Entropy \textbf{25}(2), 231 (2023).
\newblock \doi{10.3390/e25020231}.
\newblock \urlprefix\url{https://doi.org/10.3390/e25020231}

\bibitem{kempe2016robustness}
X.~He, D.~Kempe, \emph{Robust Influence Maximization}, in \emph{Proceedings of the 22nd ACM SIGKDD International Conference on Knowledge Discovery and Data Mining} (Association for Computing Machinery, New York, NY, USA, 2016), KDD '16, p. 885–894.
\newblock \doi{10.1145/2939672.2939760}.
\newblock \urlprefix\url{https://doi.org/10.1145/2939672.2939760}

\bibitem{ou2022identifying}
Y.~Ou, Q.~Guo, J.~Liu, Identifying spreading influence nodes for social networks.
\newblock Frontiers of Engineering Management pp. 1--30 (2022).
\newblock \doi{10.1007/s42524-022-0190-8}.
\newblock \urlprefix\url{https://link.springer.com/article/10.1007/s42524-022-0190-8}

\bibitem{salehi2015spreading}
M.~Salehi, R.~Sharma, M.~Marzolla, M.~Magnani, P.~Siyari, D.~Montesi, Spreading processes in multilayer networks.
\newblock IEEE Transactions on Network Science and Engineering \textbf{2}(2), 65--83 (2015).
\newblock \doi{10.1109/TNSE.2015.2425961}.
\newblock \urlprefix\url{https://ieeexplore.ieee.org/document/7093190}

\bibitem{zhao2013identifying}
D.~Zhao, L.~Li, S.~Li, Y.~Huo, Y.~Yang, Identifying influential spreaders in interconnected networks.
\newblock Physica Scripta \textbf{89}(1), 015203 (2013).
\newblock \doi{10.1088/0031-8949/89/01/015203}.
\newblock \urlprefix\url{https://iopscience.iop.org/article/10.1088/0031-8949/89/01/015203/pdf}

\bibitem{erlandsson2018seed}
F.~Erlandsson, P.~Br{\'o}dka, A.~Borg, \emph{Seed Selection for Information Cascade in Multilayer Networks}, in \emph{Complex Networks \& Their Applications VI: Proceedings of Complex Networks 2017 (The Sixth International Conference on Complex Networks and Their Applications)} (Springer, 2018), pp. 426--436.
\newblock \doi{10.1007/978-3-319-72150-7_35}.
\newblock \urlprefix\url{https://doi.org/10.1007/978-3-319-72150-7_35}

\bibitem{erlandsson2016finding}
F.~Erlandsson, P.~Br{\'o}dka, A.~Borg, H.~Johnson, Finding influential users in social media using association rule learning.
\newblock Entropy \textbf{18}(5), 164 (2016).
\newblock \doi{10.3390/e18050164}.
\newblock \urlprefix\url{https://www.mdpi.com/1099-4300/18/5/164}

\bibitem{chen2020maximizing}
Y.~Chen, W.~Wang, J.~Feng, Y.~Lu, X.~Gong, Maximizing multiple influences and fair seed allocation on multilayer social networks.
\newblock Plos one \textbf{15}(3), e0229201 (2020).
\newblock \doi{10.1371/journal.pone.0229201}.
\newblock \urlprefix\url{https://doi.org/10.1371/journal.pone.0229201}

\bibitem{wang2022mfearim}
S.~Wang, X.~Tan, Determining seeds with robust influential ability from multi-layer networks: A multi-factorial evolutionary approach.
\newblock Knowledge-Based Systems \textbf{246}, 108697 (2022).
\newblock \doi{https://doi.org/10.1016/j.knosys.2022.108697}.
\newblock \urlprefix\url{https://www.sciencedirect.com/science/article/pii/S0950705122003215}

\bibitem{wang2022marimmulti}
S.~Wang, X.~Tan, Solving the robust influence maximization problem on multi-layer networks via a memetic algorithm.
\newblock Applied Soft Computing \textbf{121}, 108750 (2022).
\newblock \doi{https://doi.org/10.1016/j.asoc.2022.108750}.
\newblock \urlprefix\url{https://www.sciencedirect.com/science/article/pii/S1568494622001892}

\bibitem{qiang2018rspn3}
Q.~He, X.~Wang, M.~Huang, J.~Lv, L.~Ma, Heuristics-based influence maximization for opinion formation in social networks.
\newblock Applied Soft Computing \textbf{66}, 360--369 (2018).
\newblock \doi{https://doi.org/10.1016/j.asoc.2018.02.016}.
\newblock \urlprefix\url{https://www.sciencedirect.com/science/article/pii/S1568494618300759}

\bibitem{singh2019lapso}
S.S. Singh, A.~Kumar, K.~Singh, B.~Biswas, Lapso-im: A learning-based influence maximization approach for social networks.
\newblock Appl. Soft Comput. \textbf{82}(C) (2019).
\newblock \doi{10.1016/j.asoc.2019.105554}.
\newblock \urlprefix\url{https://doi.org/10.1016/j.asoc.2019.105554}

\bibitem{Iacca2021imea}
G.~Iacca, K.~Konotopska, D.~Bucur, A.~Tonda, {An evolutionary framework for maximizing influence propagation in social networks}.
\newblock Software Impacts \textbf{9} (2021).
\newblock \doi{10.1016/j.simpa.2021.100107}.
\newblock \urlprefix\url{https://www.softwareimpacts.com/article/S2665-9638(21)00041-5/fulltext}

\bibitem{XIE2021CIMMIC}
X.~Xie, J.~Li, Y.~Sheng, W.~Wang, W.~Yang, Competitive influence maximization considering inactive nodes and community homophily.
\newblock Knowledge-Based Systems \textbf{233}, 107497 (2021).
\newblock \doi{https://doi.org/10.1016/j.knosys.2021.107497}.
\newblock \urlprefix\url{https://www.sciencedirect.com/science/article/pii/S0950705121007590}

\bibitem{JABARILOTF2022GDA}
J.~{Jabari Lotf}, M.~{Abdollahi Azgomi}, M.R. {Ebrahimi Dishabi}, An improved influence maximization method for social networks based on genetic algorithm.
\newblock Physica A: Statistical Mechanics and its Applications \textbf{586}, 126480 (2022).
\newblock \doi{https://doi.org/10.1016/j.physa.2021.126480}.
\newblock \urlprefix\url{https://www.sciencedirect.com/science/article/pii/S0378437121007536}

\bibitem{lu2020nsgaii}
Q.~Lu, Z.~Bu, Y.~Wang, \emph{A Multiobjective Evolutionary Approach for Influence Maximization in Multilayer Networks}, in \emph{Proceedings of the 2020 6th International Conference on Computing and Artificial Intelligence} (Association for Computing Machinery, New York, NY, USA, 2020), ICCAI '20, p. 431–438.
\newblock \doi{10.1145/3404555.3404568}.
\newblock \urlprefix\url{https://doi.org/10.1145/3404555.3404568}

\bibitem{Chen2009DegreeDiscount}
W.~Chen, Y.~Wang, S.~Yang, \emph{Efficient influence maximization in social networks}, in \emph{Proceedings of the 15th ACM SIGKDD International Conference on Knowledge Discovery and Data Mining} (Association for Computing Machinery, New York, NY, USA, 2009), KDD '09, p. 199–208.
\newblock \doi{10.1145/1557019.1557047}.
\newblock \urlprefix\url{https://doi.org/10.1145/1557019.1557047}

\bibitem{page1999pagerank}
L.~Page, S.~Brin, R.~Motwani, T.~Winograd, The pagerank citation ranking: Bringing order to the web.
\newblock Technical Report 1999-66, Stanford InfoLab (1999).
\newblock \urlprefix\url{http://ilpubs.stanford.edu:8090/422/}.
\newblock Previous number = SIDL-WP-1999-0120

\bibitem{Venkatakrishna2022CIM}
V.{\relax Rao}. K, M.~Katukuri, M.~Jagarapu, {CIM: clique-based heuristic for finding influential nodes in multilayer networks}.
\newblock Applied Intelligence \textbf{52}(5), 5173--5184 (2022).
\newblock \doi{10.1007/s10489-021-02656-0}.
\newblock \urlprefix\url{https://doi.org/10.1007/s10489-021-02656-0}

\bibitem{Venkatakrishna2023KppShell}
V.R. K., C.R. Chowdary, {K++ Shell: Influence maximization in multilayer networks using community detection}.
\newblock Computer Networks \textbf{234}, 109916 (2023).
\newblock \doi{10.1016/j.comnet.2023.109916}.
\newblock \urlprefix\url{https://www.sciencedirect.com/science/article/pii/S1389128623003614}

\bibitem{Narayanam2011spins}
R.~Narayanam, Y.~Narahari, A shapley value-based approach to discover influential nodes in social networks.
\newblock IEEE Transactions on Automation Science and Engineering \textbf{8}(1), 130--147 (2011).
\newblock \doi{10.1109/TASE.2010.2052042}.
\newblock \urlprefix\url{https://ieeexplore.ieee.org/document/5499450}

\bibitem{kuhnle2018knapsackseeding}
A.~Kuhnle, M.A. Alim, X.~Li, H.~Zhang, M.T. Thai, Multiplex influence maximization in online social networks with heterogeneous diffusion models.
\newblock IEEE Transactions on Computational Social Systems \textbf{5}(2), 418--429 (2018).
\newblock \doi{10.1109/TCSS.2018.2813262}.
\newblock \urlprefix\url{https://ieeexplore.ieee.org/document/8332491}

\bibitem{li2019disco}
H.~Li, M.~Xu, S.S. Bhowmick, C.~Sun, Z.~Jiang, J.~Cui.
\newblock Disco: Influence maximization meets network embedding and deep learning (2019).
\newblock \doi{10.48550/arXiv.1906.07378}.
\newblock \urlprefix\url{https://doi.org/10.48550/arXiv.1906.07378}

\bibitem{shai2007kshell}
S.~Carmi, S.~Havlin, S.~Kirkpatrick, Y.~Shavitt, E.~Shir, A model of internet topology using <i>k</i>-shell decomposition.
\newblock Proceedings of the National Academy of Sciences \textbf{104}(27), 11150--11154 (2007).
\newblock \doi{10.1073/pnas.0701175104}.
\newblock \urlprefix\url{https://www.pnas.org/doi/abs/10.1073/pnas.0701175104}

\bibitem{SUN20211commkshell}
P.G. Sun, Q.~Miao, S.~Staab, Community-based k-shell decomposition for identifying influential spreaders.
\newblock Pattern Recognition \textbf{120}, 108130 (2021).
\newblock \doi{https://doi.org/10.1016/j.patcog.2021.108130}.
\newblock \urlprefix\url{https://www.sciencedirect.com/science/article/pii/S0031320321003174}

\bibitem{Venkatakrishna2022CBIM}
K.~{Venkatakrishna Rao}, C.R. Chowdary, Cbim: Community-based influence maximization in multilayer networks.
\newblock Information Sciences \textbf{609}, 578--594 (2022).
\newblock \doi{10.1016/j.ins.2022.07.103}.
\newblock \urlprefix\url{https://www.sciencedirect.com/science/article/pii/S0020025522007927}

\bibitem{Chen2020imcs}
J.~Chen, J.~Liu, Research on a novel influence maximization algorithm based on community structure.
\newblock Journal of Physics: Conference Series \textbf{1631}(1), 012064 (2020).
\newblock \doi{10.1088/1742-6596/1631/1/012064}.
\newblock \urlprefix\url{https://dx.doi.org/10.1088/1742-6596/1631/1/012064}

\bibitem{Kumar2021imcshindex}
S.~Kumar, L.~Singhla, K.~Jindal, K.~Grover, B.S. Panda, {IM-ELPR: Influence maximization in social networks using label propagation based community structure}.
\newblock Appl. Intell. \textbf{51}(11), 7647--7665 (2021).
\newblock \doi{10.1007/s10489-021-02266-w}.
\newblock \urlprefix\url{https://doi.org/10.1007/s10489-021-02266-w}

\bibitem{michalski2013convince}
R.~Michalski, P.~Kazienko, J.~Jankowski, \emph{Convince a dozen more and succeed--the influence in multi-layered social networks}, in \emph{2013 International Conference on Signal-Image Technology \& Internet-Based Systems} (IEEE, 2013), pp. 499--505.
\newblock \doi{10.1109/SITIS.2013.85}.
\newblock \urlprefix\url{https://ieeexplore.ieee.org/document/6727235}

\bibitem{zhong2022mltm}
Y.D. Zhong, V.~Srivastava, N.E. Leonard, Influence spread in the heterogeneous multiplex linear threshold model.
\newblock IEEE Transactions on Control of Network Systems \textbf{9}(3), 1080--1091 (2022).
\newblock \doi{10.1109/TCNS.2021.3088782}.
\newblock \urlprefix\url{https://ieeexplore.ieee.org/document/9454288}

\bibitem{ziolo2022modeling}
M.~Zio{\l}o, P.~Br{\'o}dka, A.~Spoz, J.~Jankowski, Modeling the impact of external influence on green behaviour spreading in multilayer financial networks.
\newblock 2022 IEEE 9th International Conference on Data Science and Advanced Analytics (DSAA) pp. 1--10 (2022).
\newblock \doi{10.1109/DSAA54385.2022.10032397}.
\newblock \urlprefix\url{https://ieeexplore.ieee.org/document/10032397}

\bibitem{dickison2016multilayer}
M.E. Dickison, M.~Magnani, L.~Rossi, \emph{Multilayer social networks} (Cambridge University Press, Cambridge, England, UK, 2016).
\newblock \doi{10.1017/CBO9781139941907}.
\newblock \urlprefix\url{https://doi.org/10.1017/CBO9781139941907}

\bibitem{kivela2014multilayer}
M.~Kivel{\"a}, A.~Arenas, M.~Barthelemy, J.P. Gleeson, Y.~Moreno, M.A. Porter, Multilayer networks.
\newblock Journal of Complex Networks \textbf{2}(3), 203--271 (2014).
\newblock \doi{10.1093/comnet/cnu016}.
\newblock \urlprefix\url{https://doi.org/10.1093/comnet/cnu016}

\bibitem{granovetter1978threshold}
M.~Granovetter, Threshold models of collective behavior.
\newblock American journal of sociology \textbf{83}(6), 1420--1443 (1978).
\newblock \urlprefix\url{http://www.jstor.org/stable/2778111}

\bibitem{magnani2011ml}
M.~Magnani, L.~Rossi, \emph{The ML-Model for Multi-layer Social Networks}, in \emph{2011 International Conference on Advances in Social Networks Analysis and Mining} (2011), pp. 5--12.
\newblock \doi{10.1109/ASONAM.2011.114}.
\newblock \urlprefix\url{https://ieeexplore.ieee.org/document/5992579}

\bibitem{dice1945similarity}
L.R. Dice.
\newblock {Measures of the Amount of Ecologic Association Between Species} (1945).
\newblock \doi{doi.org/10.2307/1932409}.
\newblock \urlprefix\url{https://www.jstor.org/stable/1932409}

\bibitem{katz1953centrality}
L.~Katz, {A new status index derived from sociometric analysis}.
\newblock Psychometrika \textbf{18}(1), 39--43 (1953).
\newblock \doi{10.1007/BF02289026}.
\newblock \urlprefix\url{https://doi.org/10.1007/BF02289026}

\bibitem{magnani2021analysis}
M.~Magnani, L.~Rossi, D.~Vega, Analysis of multiplex social networks with r.
\newblock Journal of Statistical Software \textbf{98}, 1--30 (2021)

\bibitem{dedomenico2015arxiv}
M.~De~Domenico, A.~Lancichinetti, A.~Arenas, M.~Rosvall, Identifying modular flows on multilayer networks reveals highly overlapping organization in interconnected systems.
\newblock Phys. Rev. X \textbf{5}, 011027 (2015).
\newblock \doi{10.1103/PhysRevX.5.011027}.
\newblock \urlprefix\url{https://link.aps.org/doi/10.1103/PhysRevX.5.011027}

\bibitem{rossi2015aucs}
L.~Rossi, M.~Magnani, Towards effective visual analytics on multiplex and multilayer networks.
\newblock Chaos, Solitons \& Fractals \textbf{72}, 68--76 (2015).
\newblock \doi{10.1016/j.chaos.2014.12.022}.
\newblock \urlprefix\url{https://www.sciencedirect.com/science/article/pii/S0960077914002422}.
\newblock Multiplex Networks: Structure, Dynamics and Applications

\bibitem{coleman1957ckmp}
J.~Coleman, E.~Katz, H.~Menzel, The diffusion of an innovation among physicians.
\newblock Sociometry \textbf{20}(4), 253--270 (1957).
\newblock \urlprefix\url{http://www.jstor.org/stable/2785979}

\bibitem{cardillo2013eutransportation}
A.~Cardillo, J.~G{\ifmmode\acute{o}\else\'{o}\fi}mez-Garde{\ifmmode\tilde{n}\else\~{n}\fi}es, M.~Zanin, M.~Romance, D.~Papo, F.d. Pozo, S.~Boccaletti, {Emergence of network features from multiplexity}.
\newblock Scientific Reports \textbf{3}(1344), 1--6 (2013).
\newblock \doi{10.1038/srep01344}.
\newblock \urlprefix\url{https://www.nature.com/articles/srep01344}

\bibitem{snijders2006lazega}
T.A.B. Snijders, P.E. Pattison, G.L. Robins, M.S. Handcock, New specifications for exponential random graph models.
\newblock Sociological Methodology \textbf{36}(1), 99--153 (2006).
\newblock \doi{10.1111/j.1467-9531.2006.00176.x}.
\newblock \urlprefix\url{https://doi.org/10.1111/j.1467-9531.2006.00176.x}

\bibitem{jankowski2024timik}
J.~Jankowski, R.~Michalski, P.~Br\'{o}dka, A multilayer network dataset of interaction and influence spreading in a virtual world.
\newblock Scientific Data \textbf{4}(1), 170144 (2017).
\newblock \doi{10.1038/sdata.2017.144}.
\newblock \urlprefix\url{https://doi.org/10.1038/sdata.2017.144}

\bibitem{er-model}
P.~Erd{\H{o}}s, A.~R{\'e}nyi, et~al., On the evolution of random graphs.
\newblock Publ. math. inst. hung. acad. sci \textbf{5}(1), 17--60 (1960).
\newblock \urlprefix\url{https://static.renyi.hu/~p_erdos/1960-10.pdf}

\bibitem{sf-model}
B.~Bollobas, C.~Borgs, J.~Chayes, O.~Riordan, \emph{Directed Scale-Free Graphs}, in \emph{Proceedings of the 14th Annual ACM-SIAM Symposium on Discrete Algorithms (SODA)} (2003), pp. 132--139.
\newblock \urlprefix\url{https://www.microsoft.com/en-us/research/publication/directed-scale-free-graphs/}

\bibitem{hagberg2008networkx}
A.A. Hagberg, D.A. Schult, P.J. Swart, \emph{Exploring Network Structure, Dynamics, and Function using NetworkX}, in \emph{Proceedings of the 7th Python in Science Conference}, ed. by G.~Varoquaux, T.~Vaught, J.~Millman (Pasadena, CA USA, 2008), pp. 11 -- 15.
\newblock \urlprefix\url{https://aric.hagberg.org/papers/hagberg-2008-exploring.pdf}

\bibitem{czuba2022networkdiffusion}
M.~Czuba, P.~Br\'{o}dka, \emph{Simulating Spreading of Multiple Interacting Processes in Complex Networks}, in \emph{2022 IEEE 9th International Conference on Data Science and Advanced Analytics (DSAA)} (IEEE, Shenzhen, China, 2022), pp. 1--10.
\newblock \doi{10.1109/DSAA54385.2022.10032425}.
\newblock \urlprefix\url{https://ieeexplore.ieee.org/abstract/document/10032425}

\bibitem{czuba2024networkdiffusion}
M.~Czuba, M.~Nurek, D.~Serwata, Y.X. Qi, M.~Jia, K.~Musial, R.~Michalski, P.~Br{\'o}dka, Network diffusion framework to simulate spreading processes in complex networks.
\newblock Big Data Mining And Analytics \textbf{7}(3), 637--654 (2024).
\newblock \doi{10.26599/BDMA.2024.9020010}.
\newblock \urlprefix\url{https://doi.org/10.26599/BDMA.2024.9020010}

\bibitem{conover1999practical}
W.J. Conover, \emph{{Practical Nonparametric Statistics}} (Wiley, Hoboken, NJ, USA, 1999).
\newblock \urlprefix\url{https://www.wiley.com/en-us/Practical+Nonparametric+Statistics%2C+3rd+Edition-p-9780471160687}

\bibitem{hassan2020operational}
B.A. Hassan, T.A. Rashid, Operational framework for recent advances in backtracking search optimisation algorithm: A systematic review and performance evaluation.
\newblock Applied Mathematics and Computation \textbf{370}, 124919 (2020).
\newblock \doi{10.1016/j.amc.2019.124919}.
\newblock \urlprefix\url{https://www.sciencedirect.com/science/article/pii/S0096300319309117}

\bibitem{liu2025tnnls_2023_3321767}
Y.~Liu, X.~Wang, X.~Wang, Z.~Wang, J.~Kurths, Diffusion source inference for large-scale complex networks based on network percolation.
\newblock IEEE Transactions on Neural Networks and Learning Systems \textbf{36}(1), 1453--1466 (2025).
\newblock \doi{10.1109/TNNLS.2023.3321767}.
\newblock \urlprefix\url{https://doi.org/10.1109/TNNLS.2023.3321767}

\bibitem{liu2024tnet_2024_3382546}
Y.~Liu, X.~Wang, X.~Wang, Z.~Wang, Fast outbreak sense and effective source inference via minimum observer set.
\newblock IEEE/ACM Transactions on Networking \textbf{32}(4), 3111--3125 (2024).
\newblock \doi{10.1109/TNET.2024.3382546}.
\newblock \urlprefix\url{https://doi.org/10.1109/TNET.2024.3382546}

\bibitem{liu2024tifs_2023_3338423}
Y.~Liu, G.~Liang, X.~Wang, P.~Zhu, Z.~Wang, Diffusion containment in complex networks through collective influence of connections.
\newblock IEEE Transactions on Information Forensics and Security \textbf{19}, 1510--1524 (2024).
\newblock \doi{10.1109/TIFS.2023.3338423}.
\newblock \urlprefix\url{https://doi.org/10.1109/TIFS.2023.3338423}

\bibitem{liu2024tnse_2024_3388994}
Y.~Liu, Y.~Zhong, X.~Li, P.~Zhu, Z.~Wang, Vital nodes identification via evolutionary algorithm with percolation optimization in complex networks.
\newblock IEEE Transactions on Network Science and Engineering \textbf{11}(4), 3838--3850 (2024).
\newblock \doi{10.1109/TNSE.2024.3388994}.
\newblock \urlprefix\url{https://doi.org/10.1109/TNSE.2024.3388994}

\end{thebibliography}

\end{document}